# Relating the thermal properties of a micro pulsating heat pipe to the internal flow characteristics via experiments, image recognition of flow patterns and heat transfer simulations

Chihiro Kamijima, Yuta Yoshimoto\*, Yutaro Abe, Shu Takagi, Ikuya Kinefuchi\*

*Department of Mechanical Engineering, The University of Tokyo, 7-3-1 Hongo, Bunkyo-ku, Tokyo 113-8656, Japan*

ABSTRACT

We investigate the relationship between the thermal properties of a micro pulsating heat pipe (MPHP) and the internal flow characteristics via measurements of effective thermal conductivities, flow visualization followed by image recognition of the flow patterns, and heat transfer simulations employing the extracted flow patterns. The MPHP consists of an eleven-turn closed-loop of a meandering square microchannel with a hydraulic diameter of 350 μm engraved on a silicon substrate, which is covered with a transparent glass plate to allow internal flow visualization. The MPHP charged with Fluorinert FC-72 tends to exhibit higher effective thermal conductivities for the coolant temperature of $T_c = 40$ °C compared to $T_c = 20$ °C, and provides the highest effective thermal conductivity of about 700 W/(m·K) for $T_c = 40$ °C and a filling ratio of 48%. Interestingly, we observe two different self-oscillation modes having different thermal conductivities, even for identical heat input rates. This tendency indicates a hysteresis of the effective thermal conductivity, which originates from the difference in the heat input rates at which the MPHP falls into and recovers from dryout. Subsequently, semantic segmentation-based image recognition is applied to the recorded flow images to identify the flow characteristics, successfully extracting four different flow patterns involving liquid slugs, liquid films, dry walls, and rapid-boiling regions. The image recognition results indicate that high effective thermal conductivities of the MPHP relate to stable self-oscillations with large amplitudes and high frequencies, along with long and thin liquid films beneficial for latent heat

---

\* Corresponding authors
Email addresses: yyoshimoto@fel.t.u-tokyo.ac.jp (Y. Yoshimoto), kine@fel.t.u-tokyo.ac.jp (I. Kinefuchi)

transfer. Finally, we perform numerical simulations of latent/sensible heat transfer via vapor plugs and of sensible heat transfer via liquid slugs using the extracted flow patterns and measured channel temperatures as inputs. We find that latent heat transfer via liquid films accounts for a considerable portion of the overall heat transfer, while the sensible heat transfer via liquid slugs is much less significant.

Disclosure statement

The authors declare no conflict of interest.

Nomenclature

| | |
|---|---|
| $a$ | side length of the microchannel |
| $c_{p,l}$ | specific heat of liquid |
| $d$ | distance between thermocouples 1 and 2 |
| $\dot{H}_{lat,i}$ | inflow rate of enthalpy to vapor plug $i$ |
| $h_{lv}$ | specific heat of vaporization |
| $h_v$ | specific vapor enthalpy |
| $L_i$ | length of liquid slug $i$ |
| $m''$ | mass flux to vapor due to phase change of liquid film |
| $n_i$ | molar number of vapor plug $i$ |
| $Nu$ | Nusselt number |
| $P_i$ | pressure of vapor plug $i$ |
| $\dot{Q}_H$ | heat input rate |
| $\dot{Q}_P$ | heat transfer rate through the MPHP |
| $\dot{Q}_W$ | heat transfer rate through interconnections |
| $\dot{Q}_{sen,i}$ | sensible heat transfer rate to vapor plug $i$ |
| $\dot{Q}_{v,i}^{cnd}$ | heat transfer rate to the condenser via vapor plug $i$ |
| $\dot{Q}_{v}^{cnd}$ | heat transfer rate to the condenser via all vapor plugs |
| $Q_{v}^{cnd}$ | cumulative heat transferred to the condenser via all vapor plugs |
| $\dot{Q}_{wall}^{cnd}$ | heat transfer rate to the condenser via channel wall |
| $\dot{Q}_{l,i}^{cnd}$ | heat transfer rate to the condenser via liquid slug $i$ |
| $\dot{Q}_{l}^{cnd}$ | heat transfer rate to the condenser via all liquid slugs |
| $q''_{sen}$ | sensible heat flux from liquid film/channel wall to vapor |
| $q''_{lat}$ | latent heat flux from liquid film to vapor |
| $q''_w$ | wall heat flux |
| $R$ | universal gas constant |
| $R_P$ | thermal resistance of the MPHP |
| $R_W$ | thermal resistance between the evaporator and the environment |
| $r$ | coordinate perpendicular to the flow direction |
| $S_{PHP}$ | cross-sectional area of the MPHP |
| $S_{si}$ | cross-sectional area of the silicon substrate |
| $T_c$ | coolant temperature |
| $T_1$ | temperature measured by thermocouple 1 |
| $T_2$ | temperature measured by thermocouple 2 |
| $T_L$ | condenser temperature |
| $T_H$ | evaporator temperature |
| $T_0$ | environmental temperature |
| $T_{v,i}$ | temperature of vapor plug $i$ |
| $T_{sur}$ | surface temperature of channel wall or liquid film |
| $T_w$ | wall temperature |
| $T_{sat}$ | saturation temperature |
| $T_{l,i}$ | temperature of liquid slug $i$ |
| $U_i$ | internal energy of vapor plug $i$ |
| $V_i$ | volume of vapor plug $i$ |
| $x$ | coordinate in the flow direction |
| $x_{g,i}$ | center-of-mass of liquid slug $i$ |
| $X_i$ | coordinate fixed at liquid slug $i$ |
| $z_i$ | compressibility factor of vapor plug $i$ |

*Greek symbols*

| | |
|---|---|
| $\delta$ | liquid film thickness |
| $\varphi$ | filling ratio of working fluid |
| $\lambda_{eff}$ | effective thermal conductivity of the MPHP |
| $\lambda_v$ | thermal conductivity of vapor |
| $\lambda_l$ | thermal conductivity of liquid |
| $\lambda_{si}$ | thermal conductivity of silicon substrate |
| $\rho_l$ | liquid density |

1. Introduction

With the integration and miniaturization of electronic devices in recent years, demand for high-performance, small heat-transfer devices has arisen to cope with the large heat generation and high temperature of IC chips. Heat pipes [1] are one of the most widely used devices, owing to their large thermal conductivities. Conventional heat pipes use wicks to circulate working fluids from a condenser to an evaporator via the capillary force. However, it is difficult to minimize such heat pipes to the microscale because of capillary limits and difficulties with manufacturing complex structures, including many wicks. To address these issues, pulsating heat pipes (PHPs), first introduced by Akachi et al. [2], have attracted much attention due to their ease of miniaturization, simple structures, and excellent thermal properties. PHPs consist of meandering flow channels between cooling and heating sections. When PHPs are partially filled with working fluid, multiple liquid slugs and vapor plugs form in the flow channels. PHPs utilize self-oscillating slug flows caused by the pressure differences of vapor plugs for effective heat transfer.

There are many parameters that affect the complex and chaotic self-oscillations of working fluid; among these are the configurations of the flow channels, including their inner diameter and number of turns, as well as operating conditions such as the filling ratio of the working fluid and the inclination angle of the PHP, and the physical properties of the working fluid [3]. Therefore, the relationship between these parameters and the thermal properties of PHPs has been widely investigated experimentally. Khandekar et al. [4] experimentally investigated the thermal properties of a PHP by changing filling ratios, input heat, and working fluid. They concluded that the filling ratio should be in the range of 25–65% to sustain effective heat transfer via the pulsating motions of liquid slugs. Yang et al. [5] investigated the influence of hydraulic diameter and inclination angle upon the thermal resistances of the PHPs. While a PHP with an inner diameter of 2 mm achieved lower thermal resistance, one with an inner diameter of 1 mm could function for larger input heat fluxes, and its thermal properties were not affected by the inclination angle. Charoensawan et al. [6] fabricated PHPs with different inner diameters and numbers of turns in flow channels to investigate the relationship between the device parameters and thermal properties. Larger heat transfer rates were attained for PHPs having larger inner diameters and more turns in the flow channel. They also concluded that the number of turns must be larger than a certain threshold for self-sustaining oscillations to commence at any inclination angle. Cui et al. [7] evaluated the thermal resistances of a PHP charged with different working fluids. Thermal properties of the PHP filled with different working fluids showed different tendencies for low heat input rates, whereas there were no significant differences in the thermal properties if the input heat was large enough.

Since Qu and Wu [8] fabricated PHPs with microscale channels engraved on silicon chips using the MEMS technique, many studies have focused upon micro-PHPs (MPHPs) because they can make the most of the PHPs' advantages: ease of miniaturization due to a wickless structure. Youn and Kim [9] fabricated an MPHP with a hydraulic diameter of 570 μm and evaluated its thermal properties using a theoretical model [10], which suggests that most heat transfer can be explained by sensible heat transfer via liquid slugs. Qu et al. [11] fabricated MPHPs with hydraulic diameters of 251 μm, 352 μm, and 394 μm; characteristic flow patterns such as nucleation boiling and circulation flow were not observed when the hydraulic diameter was 352 μm or less, resulting in larger thermal resistances compared with the PHP with a hydraulic diameter of 394 μm. Lee and Kim [12] measured the thermal resistances of MPHPs composed of square and circular channels, and concluded that square channels were advantageous because more working fluid was supplied to the evaporator by flowing along the four corners. They stated that dryout occurred when the evaporation rate exceeded the flow rate of liquid films falling to the evaporator due to gravity. Yoon and Kim [13] analyzed self-oscillating flows of an MPHP with a hydraulic diameter of 667 μm and showed that latent heat transfer via liquid films accounted for a large portion of the overall heat transfer.

Despite these studies, the heat transfer mechanisms of PHPs have yet to be adequately understood. For example, some studies [9,14–16] have suggested that sensible heat transfer via liquid slugs is significant, in contrast to the conventional heat pipes exploiting latent heat transfer. Meanwhile, there are also some studies [13,17] which have concluded that latent heat transfer is more significant than sensible heat transfer. Some studies [8,9,11–13,17–20] have visualized internal flows in transparent PHPs to elucidate the operational mechanisms therein. However, complex multiphase flows in PHPs make it difficult to extract detailed information from the flow visualization results, hindering a better understanding of flow mechanisms and heat transfer properties.

In the present study, we aim to relate the thermal properties of an MPHP to its internal flow characteristics based on measurements of effective thermal conductivities, flow visualization followed by image recognition of the flow patterns, and heat transfer simulations employing the extracted flow patterns. The MPHP consists of an eleven-turn closed-loop of a meandering square microchannel with a hydraulic diameter of 350 μm engraved on a silicon substrate, which is covered with a transparent glass plate to allow for internal flow visualization. The effective thermal conductivities of the MPHP charged with Fluorinert FC-72 are measured for different filling ratios and coolant temperatures. We also visualize the internal flows of the MPHP using a high-speed camera. Subsequently, semantic segmentation-based image recognition is applied to the recorded flow images to identify the flow characteristics, successfully extracting four different flow patterns involving liquid slugs, liquid films, dry walls, and rapid-boiling regions. Accordingly, we perform numerical simulations of latent/sensible heat transfer via vapor plugs and sensible heat transfer via liquid slugs using the extracted flow patterns and measured channel temperatures as inputs. By integrating the experiments, image recognition of the flow patterns, and heat transfer simulations, we successfully characterize the internal flow characteristics pertaining to better thermal properties of the MPHP.

## 2. Experiments
### 2.1. Fabrication of the MPHP

The eleven-turn closed-loop MPHP shown in Fig. 1 is fabricated using MEMS techniques. First, a meandering square microchannel is dry-etched on a silicon substrate, and a hole for injecting working fluid is made using an ultrasonic drill. Then, a Tempax glass plate is bonded by anodic bonding for flow visualization. A front view of the MPHP and a corresponding schematic diagram are illustrated in Figs. 1(a) and (b). The height and width of the chip are 40 mm and 20 mm. The MPHP has 11 turns and hence 22 parallel straight channels with lengths of 30.25 mm. The leftmost and rightmost channels are connected via the top channel and compose a closed-loop pulsating heat pipe. A cross-section of the MPHP is shown in Fig. 1(c). Both the silicon substrate and the glass plate have a thickness of 525 μm. Each channel has a 350 μm × 350 μm square cross-section, corresponding to a hydraulic diameter of 350 μm.

The arithmetic mean roughness of the bottom surface of the groove is below 8 nm, and the contact angle of FC-72 on the silicon substrate after RCA cleaning treatment followed by 1-day exposure to the atmosphere is about 13°. Note that the 11-turn MPHP with a hydraulic diameter of 350 μm is chosen since MPHPs having similar configurations operated successfully in the previous studies [12,20], ensuring successful observation of self-sustained oscillations of the working fluid.

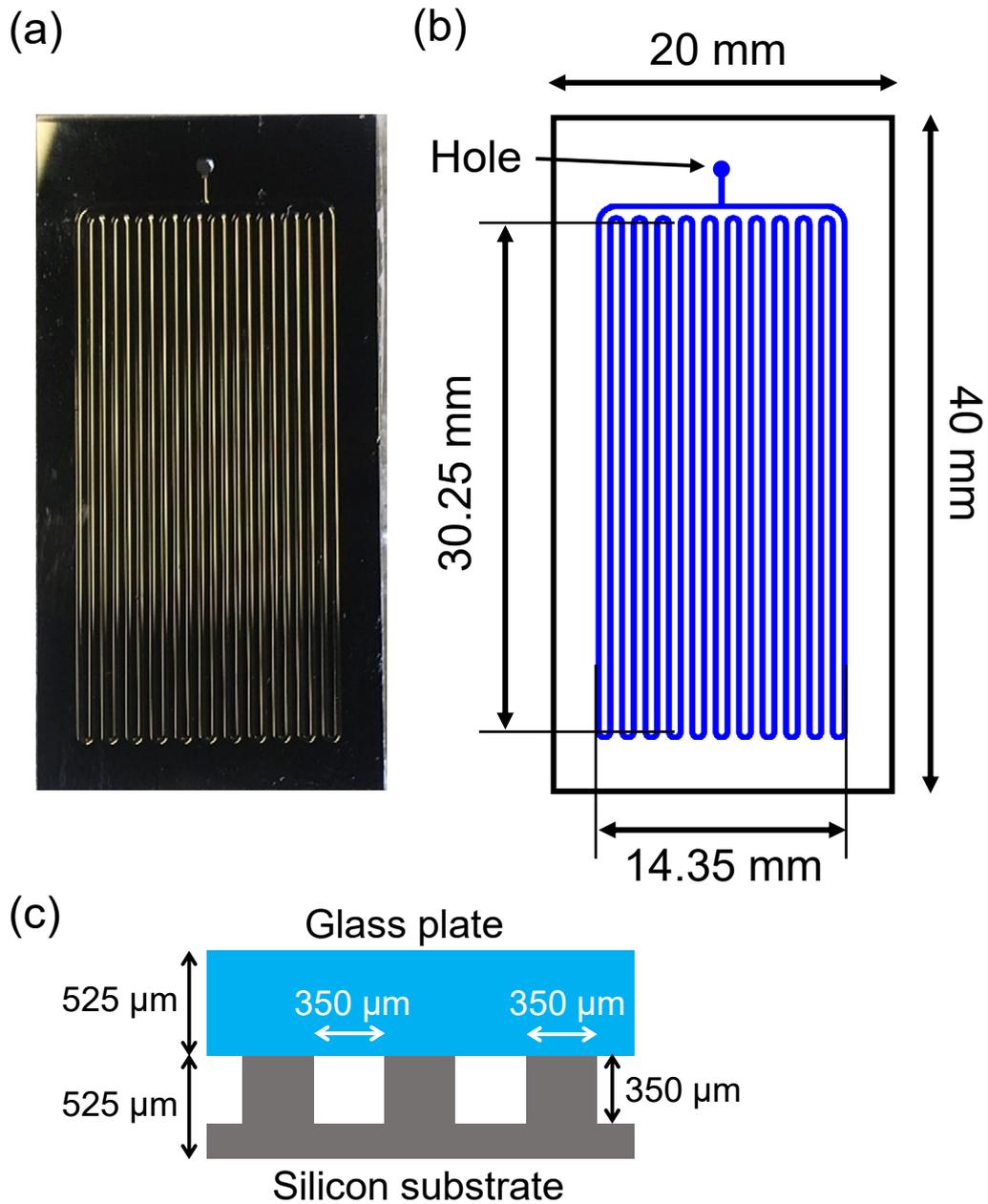

**Fig. 1.** The MPHP fabricated in this study. (a) A photograph of the front view of the MPHP. (b) A schematic diagram of the flow channels corresponding to (a). (c) A cross-section of the MPHP.

## 2.2. Experimental apparatus

An overview of our experimental system is shown in Fig. 2(a). All the experiments are conducted in a bottom-heating mode. To prevent heat transfer through the atmosphere, the MPHP is set in a vacuum chamber (Fig. 2(b)). This vacuum chamber, the channels of the MPHP, and the glass container of the working fluid are connected to a vacuum pump (Anest Iwata DVSL-100C). There is also a pipe that connects the glass container to the MPHP to charge the working fluid without exposure to the atmosphere. The experimental system has four valves (A–D), including an electromagnetic valve (SMC XSA2-22S) (A) controlled by a function generator (Agilent Technologies 33210A). Two copper blocks respectively contact the upper and lower sections of the MPHP, which correspond to the condenser and evaporator. Coolant water, with its temperature kept constant using a chiller (Yamato Scientific CLH302), flows through a hole of the condenser block during each experiment. The coolant water is circulated by a magnetic-drive pump (Iwaki MD-6K-N) at a mass flow rate of $1.48 \times 10^{-2}$ kg/s. Since it has been indicated that the condenser temperature significantly affects the thermal performance of a PHP [19], the coolant water temperature $T_c$ is set to 20°C and 40°C in the present study. An aluminum nitride heater (Sakaguchi E.H VOC WALN-5) connected to a DC power supply (Kikusui Electronics PAN70-2.5A) is glued to the back of the evaporator block. By changing the voltage of the power supply, the heat input rate to the MPHP can be changed to an arbitrary value. To reduce the contact thermal resistance, thermal grease (Sunhayato SCH-20) is applied on the contact surface of the MPHP and the copper blocks. To evaluate thermal properties of the MPHP, two k-type thermocouples (Nilaco 856101) are attached to the adiabatic section, as shown in Fig. 2(c). Additionally, a k-type thermocouple is attached to the surface of the evaporator block, and a k-type sheathed thermocouple with an outer diameter of 0.5 mm is inserted into a hole on the condenser block. These thermocouples are connected to a data logger (Graphtec GL220) and a time-series of changes in the temperature is recorded during each experiment. Internal flow patterns are also recorded with a high-speed camera (Photron SA5) in front of the chamber at a frame rate of 5,000 fps for about 4 s. The channels are lit from both sides of the camera by two lighting panels consisting of LEDs (OptoSupply OSW4XME3C1S) and diffuser plates (Shibuya Optical Co. LSD60PC-F250).

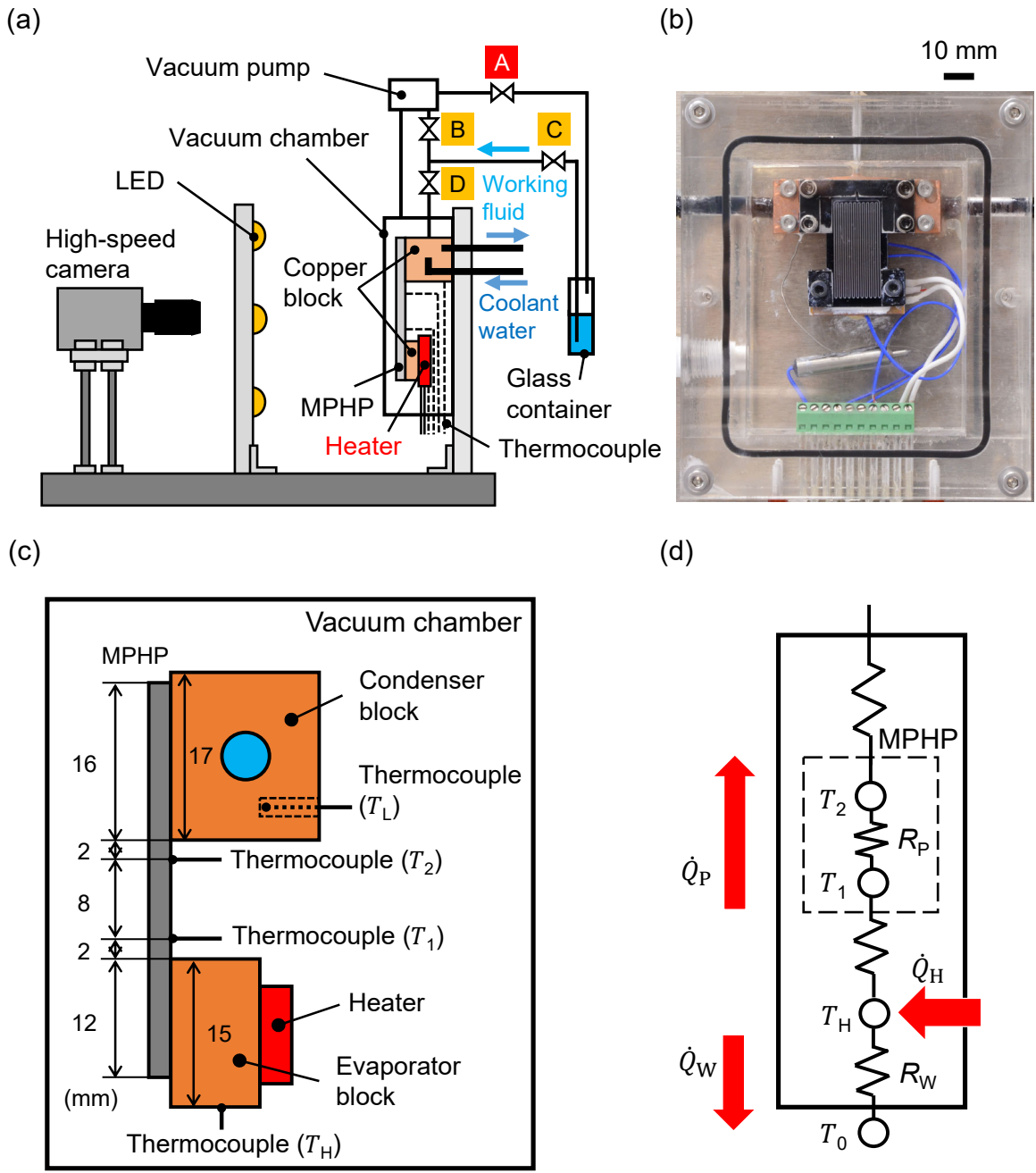

**Fig. 2.** A configuration of the experimental system. (a) A schematic diagram of the system. The two lighting panels are placed at the both sides of the vacuum chamber so that they do not disturb the high-speed camera's imaging region. (b) A picture taken from the front of the vacuum chamber. (c) Mounting positions of the MPHP and four thermocouples in the vacuum chamber. (d) The thermal equivalent circuit of the experimental system.

## 2.3. Working fluid

In this study, Fluorinert FC-72 is used as a working fluid since it is one of the typical refrigerant fluids widely used for (M)PHPs [8,11,12]. In [21], it was shown that insufficient degassing of the working fluid causes a significant decline of the thermal performance. Therefore, after charging the working fluid into the glass container and closing valves B–D (Fig. 2(a)), electromagnetic valve A is repeatedly opened for degassing for about one day. Then, valves B and D are opened to evacuate the channels of the MPHP. Subsequently, valve B is closed and valve C is opened. The pressure difference due to vapor pressure in the glass container injects the degassed working fluid into the channels.

During this process, the filling ratio $\varphi$ is monitored by the high-speed camera, and valve D is closed when this ratio reaches the prescribed value. Valve D is never opened during each experiment for sealing. The filling ratio of the working fluid varies from 39% to 63%, within which range of values the MPHPs successfully commenced operation in the previous studies [9,11,20].

## 2.4. Evaluation of thermal properties

The one-dimensional thermal equivalent circuit shown in Fig. 2(d) is used to estimate the effective heat flux through the MPHP. $T_H$, $T_1$, $T_2$, and $T_L$ correspond to the temperatures at the evaporator, two points of the adiabatic section, and the condenser, respectively (Fig. 2(c)). $T_0$ is the environmental temperature, which is always kept around 20 °C during each experiment. In this model, the heat input rate from the heater, $\dot{Q}_H$, is assumed to be transferred to the condenser through the MPHP ($\dot{Q}_P$) and to the environment through the electric wires ($\dot{Q}_W$). Then the following equations are obtained:

$$\dot{Q}_H = \dot{Q}_P + \dot{Q}_W \tag{1}$$

$$T_1 - T_2 = R_P \dot{Q}_P \tag{2}$$

$$T_H - T_0 = R_W \dot{Q}_W \tag{3}$$

where $R_P$ is the thermal resistance of the adiabatic section between the two thermocouples and $R_W$ is that between the evaporator and the environment. From Eqs. (1)–(3), $R_P$ is calculated by

$$R_P = \frac{T_1 - T_2}{\dot{Q}_H - \dfrac{T_H - T_0}{R_W}} \tag{4}$$

$R_W$ is estimated to be 103 K/W by the measurement in which a 99.994% oxygen-free copper plate of thermal conductivity 391 W/(m·K) is set between the condenser and the evaporator instead of the MPHP. $\dot{Q}_H$ is obtained by $\dot{Q}_H = VI$, where $V$ is an input voltage to the heater and $I$ is an electric current. In the present study, the effective thermal conductivity is used to evaluate the thermal performance of the MPHP, which is defined by

$$\lambda_{\text{eff}} = \frac{d}{R_P S_{\text{PHP}}} \tag{5}$$

where $S_{\text{PHP}} = 10.5$ mm$^2$ is the cross-sectional area of the MPHP and $d = 8$ mm is the distance between thermocouples 1 and 2. $S_{\text{PHP}}$ does not include the cross-sectional area of the glass plate, since the thermal conductivity of the glass plate ($\approx 1.3$ W/(m·K)) is much smaller than that of the silicon substrate ($\approx 153$ W/(m·K)) [22].

## 2.5. Experimental procedure

In each experiment, the upper limit of the heat input rate $\dot{Q}_H$ is set to be 24 W or less depending upon the evaporator temperature $T_H$. More specifically, the upper limit of $\dot{Q}_H$ is chosen such that $T_H$ does not surpass 150 ˚C to prevent the MPHP from being damaged due to excessive vapor pressure of the working fluid. Even if $T_H$ is below 150 ˚C, $\dot{Q}_H$ is raised to at most 24 W to avoid an abrupt increase in $T_H$ due to the occurrence of dryout. After the working fluid is filled into the MPHP, the heat input rate is raised to the above-mentioned upper limit to cause steady self-sustained oscillations of the working fluid. Once the system reaches a steady state, temperature measurements are conducted, along with recording of the internal flow using the high-speed camera. Subsequently, the heat input rate is reduced by 1–2 W and the same process is conducted. Unless otherwise stated, this process is repeated to obtain thermal properties against each heat input rate until the self-oscillations finally stop.

The effective thermal conductivity comes with a maximum uncertainty for the lowest heat input rate; for $T_c = 40$ ˚C, $\varphi = 39\%$, and $\dot{Q}_H = 2$ W, its uncertainty is about 19%. The uncertainty of the effective thermal conductivity decreases with increasing $\dot{Q}_H$, and does not exceed 10% for $\dot{Q}_H > 8$ W.

## 3. Thermal properties of the MPHP

### 3.1. Effective thermal conductivities

The effective thermal conductivities of the MPHP, $\lambda_{\text{eff}}$, at $T_c = 20$ °C for several filling ratios $\varphi$ are illustrated in Fig. 3(a). At any filling ratio, this conductivity improves with $\dot{Q}_H$ when $\dot{Q}_H$ is smaller than about 10 W. In this range of the heat input rate, the filling ratios of $\varphi = 39\%$ and 63% tend to yield larger thermal conductivities. However, the effective thermal conductivities for $\varphi = 39\%$, 44%, and 63% eventually decrease when $\dot{Q}_H$ exceeds 10 W, indicating the occurrence of dryout. The $\dot{Q}_H$ value corresponding to dryout is around 12 W for $\varphi = 63\%$ and around 14 W for $\varphi = 39\%$ and 44%. The effective thermal conductivity after dryout declines to about 120 W/(m·K) for $\varphi = 63\%$, which is almost the same as that of the silicon substrate itself. This suggests that self-oscillations of the working fluid hardly contribute to heat transfer. Due to the absence of dryout at least up to $\dot{Q}_H \sim 20$ W, high effective thermal conductivities for larger $\dot{Q}_H$ are measured for $\varphi = 48\%$ and 54%. Additionally, the effective thermal conductivities at $T_c = 40$ °C are shown in Fig. 3(b). Compared to the case of $T_c = 20$ °C, these conductivities tend to improve. Interestingly, for $\varphi = 48\%$, the effective thermal conductivity becomes remarkably large with the increase of $\dot{Q}_H$, reaching around 700 W/(m·K) at $\dot{Q}_H = 24$ W, the maximum value for this study. However, unlike the case at $T_c = 20$ °C, the thermal performance deteriorates for $\varphi = 54\%$ because of the onset of the dryout at $\dot{Q}_H = 14$ W, which is not observed at $T_c = 20$ °C.

Generally, filling ratios are closely related to the proportions of liquid slugs and vapor plugs, significantly affecting the thermal performances of PHPs [3]. If the filling ratio were too low, the shorter liquid slugs would benefit from the smaller wall friction force, but at the same time they would have difficulty wetting the channel wall of the evaporator continually. Meanwhile, if the filling ratio were too high, the longer liquid slugs would suffer from the larger wall friction force, thus requiring larger pressure differences of neighboring vapor plugs to drive the liquid slugs into the evaporator. We infer that the optimal filling ratio is around 50% due to the balance of these factors in the present study. Additionally, the maximum thermal conductivity of our MPHP ($\approx 700$ W/(m·K)) is consistent with that of the previous study [9], where they reported the maximum thermal conductivity of 600 W/(m·K).

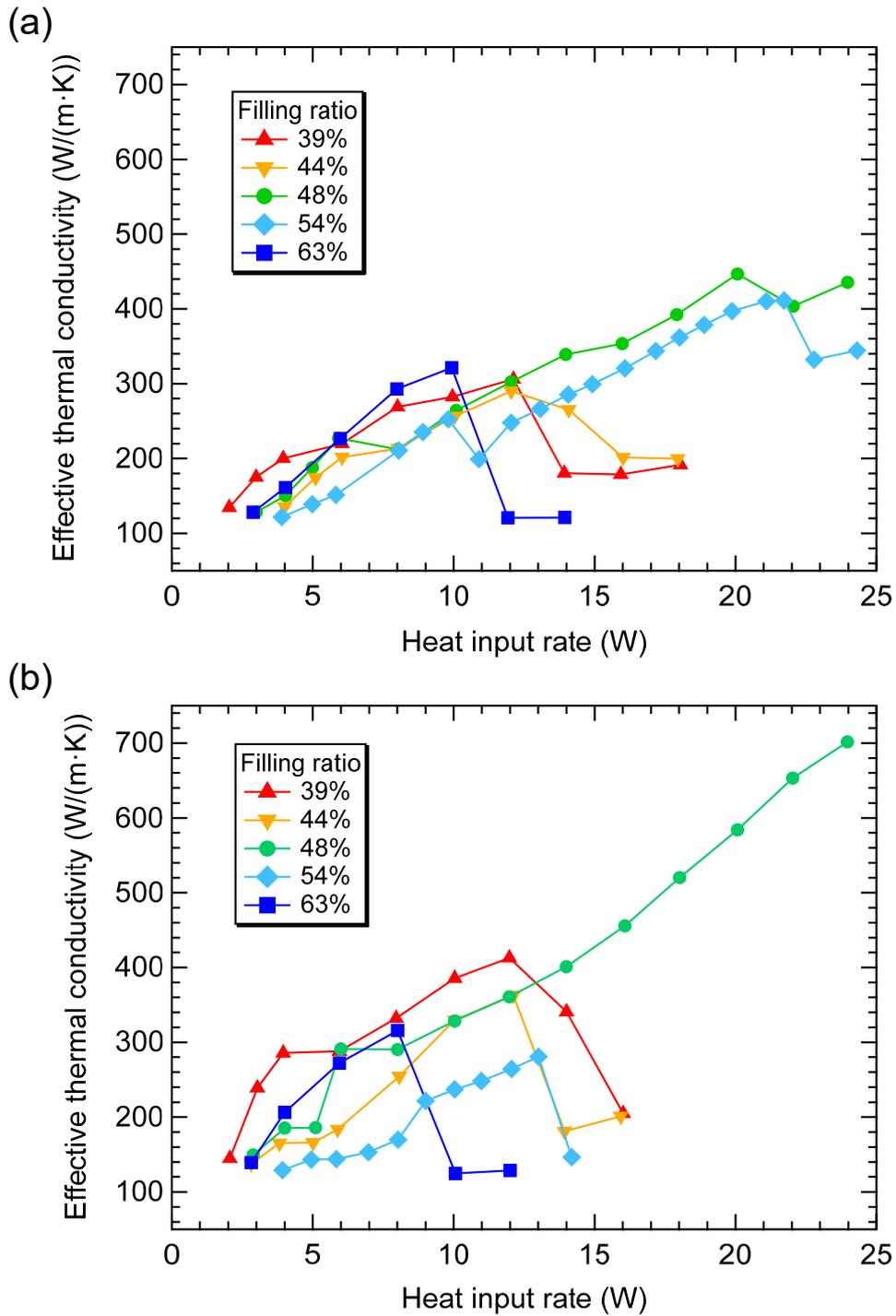

**Fig. 3.** Effective thermal conductivities for different filling ratios as a function of the heat input rate $\dot{Q}_H$ at (a) $T_c$ = 20 ˚C and (b) $T_c$ = 40 ˚C.

## 3.2. Hystereses of effective thermal conductivities

In Sec. 3.1, the way to change $\dot{Q}_H$ is consistent; once it is raised to its upper limit, it is thereafter always reduced until the self-oscillations of the working fluid stop, as described in Sec. 2.5. However, in actual situations where MPHPs are used to cool IC chips, the heat input rate would keep fluctuating. Therefore, we evaluate how the thermal properties of the MPHP depend on how $\dot{Q}_H$ is changed. The difference in the effective thermal conductivities when increasing or decreasing $\dot{Q}_H$ for $\varphi = 39\%$ at $T_c = 40$ °C is illustrated in Fig. 4(a). In case 1, $\dot{Q}_H$ is first raised until just before the evaporator temperature exceeds 150 °C ($\dot{Q}_H = 16$ W), and then reduced. In case 2, $\dot{Q}_H$ is raised from the point at which the MPHP starts up ($\dot{Q}_H = 6$ W). Although there seems to be no difference for $\dot{Q}_H \leq 12$ W, the effective thermal conductivities show different tendencies above $\dot{Q}_H = 12$ W, exhibiting a hysteresis. In case 1, the effective thermal conductivities are limited due to the dryout at $\dot{Q}_H = 16$ W before the MPHP completely recovers from the dryout at $\dot{Q}_H = 12$ W. Meanwhile, in case 2, the high effective thermal conductivities are maintained even for $\dot{Q}_H \geq 12$ W, until dryout occurs at $\dot{Q}_H = 22$ W. Additionally, the effective thermal conductivities measured for $\varphi = 39\%$ and 63% at $T_c = 20$ °C are also shown in Figs. 4(b) and (c). Again, the two distinct tendencies in cases 1 and 2 are also observed under each experimental condition, and a more significant hysteresis appears for $\varphi = 63\%$, as shown in Fig. 4(c). Although we showcase three representative conditions in Fig. 4, the hystereses of effective thermal conductivities are also observed in other conditions, and hence considered as general tendencies in the MPHP.

These results suggest that for large heat input rates, the MPHP exhibits two different self-oscillation modes that have different thermal conductivities, even for the identical heat input rates. This discrepancy originates from the difference in the heat input rates at which the MPHP falls into and recovers from dryout, as illustrated in Fig. 4(d). When $\dot{Q}_H$ is raised from a small value, the MPHP works in mode 2 with a relatively high effective thermal conductivity. However, when $\dot{Q}_H$ reaches point B, the oscillation mode transitions to mode 1 with a sudden decline of the effective thermal conductivity. Once this transition occurs, the dryout situation continues until $\dot{Q}_H$ is lowered to point A, forming a hysteresis of the effective thermal conductivity. This provides a reasonable explanation for why the MPHP has prominent thermal performances under particular conditions, such as $\varphi = 48\%$ at $T_c = 40$ °C shown in Fig. 3(b). Under this condition, the MPHP works in mode 2 without experiencing dryout throughout the experiment, leading to larger effective thermal conductivities. Therefore, from a practical viewpoint, it would be desirable for MPHPs to operate with $\dot{Q}_\mathrm{H}$ below a certain threshold to prevent the dryout leading to a hysteresis of the effective thermal conductivities.

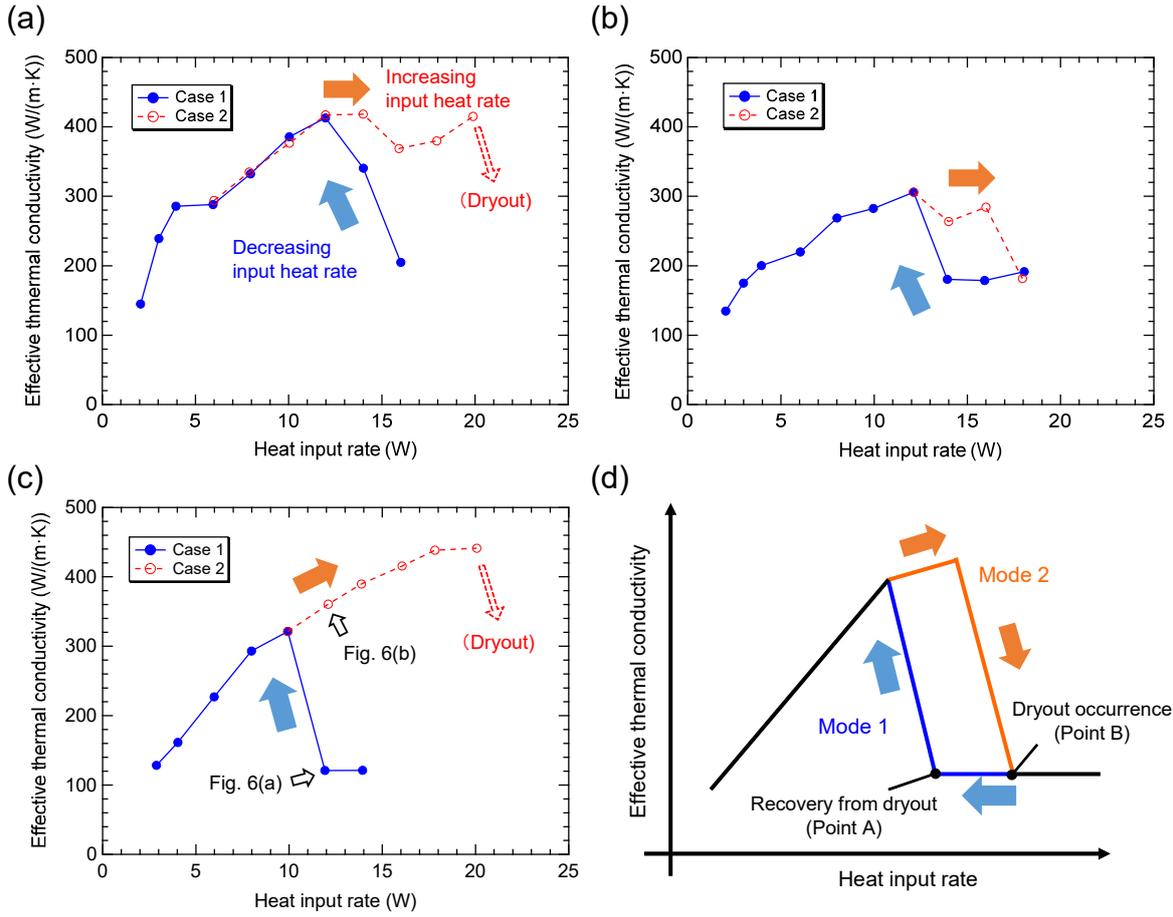

**Fig. 4.** Thermal conductivity hystereses for (a) $\varphi = 39\%$ at $T_\mathrm{c} = 40$ °C, (b) $\varphi = 39\%$ at $T_\mathrm{c} = 20$ °C, and (c) $\varphi = 63\%$ at $T_\mathrm{c} = 20$ °C. Cases 1 and 2 correspond to decreasing and increasing $\dot{Q}_\mathrm{H}$. "Dryout" in (a) and (c) means that a significant decline of the effective thermal conductivity is observed, but the complete measurement is not performed after dryout occurs because the evaporator temperature soars to over 150 °C. (d) A schematic diagram of the hysteresis. Modes 1 and 2 correspond to self-oscillation modes having small and large thermal conductivities.

## 4. Flow visualization

### 4.1. *Internal flows recorded by the high-speed camera*

In the present study, the internal flows of the MPHP are recorded using the high-speed camera. An example of a recorded internal flow image for $\varphi = 48\%$, $T_c = 40\ °C$, and $\dot{Q}_H = 20$ W is illustrated in Fig. 5, showing the representative flow behavior in which liquid slugs and vapor plugs appear alternately along the flow channel. More precisely, the regions with two vertical straight bright edges existing around the condenser correspond to liquid slugs. Relatively dark regions before and after these slugs are liquid films formed on the channel wall. The regions that look similar to liquid slugs but exist in the evaporator are dry walls without liquid films. Both liquid films and dry walls belong to vapor plugs, but they are distinguished according to whether or not the channel walls are wet. In addition to these flow patterns, the regions in which brightness changes irregularly are observed in the evaporator, corresponding to rapid-boiling.

For $\varphi = 63\%$ at $T_c = 20\ °C$, a remarkable hysteresis of the effective thermal conductivities is observed, as shown in Fig. 4(c). This is due to distinct self-oscillation modes illustrated in Fig. 6. The amplitude of the self-oscillation corresponding to mode 1 at $\dot{Q}_H = 12$ W (Fig. 6(a)) is quite small and the liquid slugs are almost "stopping". This mode causes a large drop in the effective thermal conductivity to 120 W/(m·K), meaning that the working fluid barely contributes to heat transfer. In contrast, as shown in Fig. 6(b), the working fluid self-oscillates at $\dot{Q}_H = 12$ W under mode 2, leaving a significant amount of liquid film on the channel wall. This difference in the flow behavior results in a large difference in the effective thermal conductivities shown in Fig. 4(c).

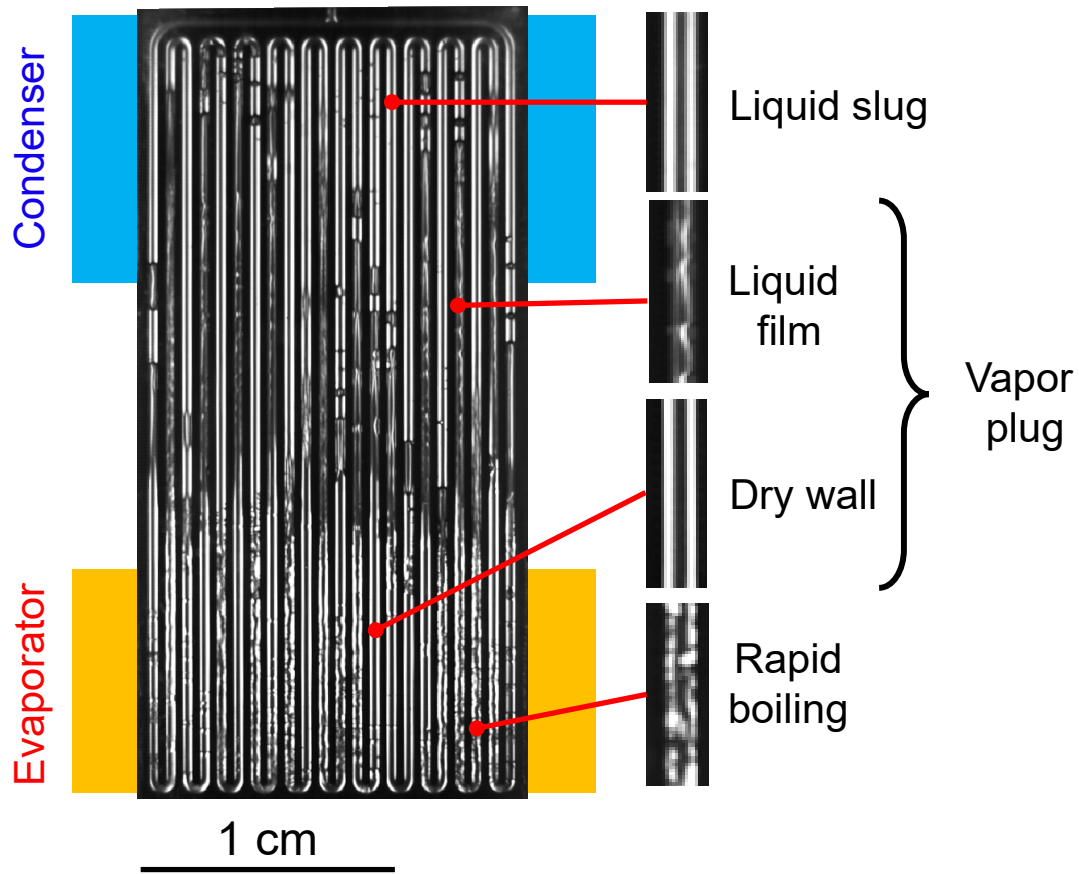

**Fig. 5.** An example of an internal flow image taken with a high-speed camera for $\varphi = 48\%$, $T_c = 40\ °C$, and $\dot{Q}_H = 20\ W$. The upper blue and lower orange sections correspond to the condenser and evaporator, respectively. Four types of flow patterns are identified, including liquid slugs, liquid films, dry walls, and rapid-boiling regions. Liquid films and dry walls are distinguished according to whether or not the channel walls are wet, both belonging to vapor plugs.

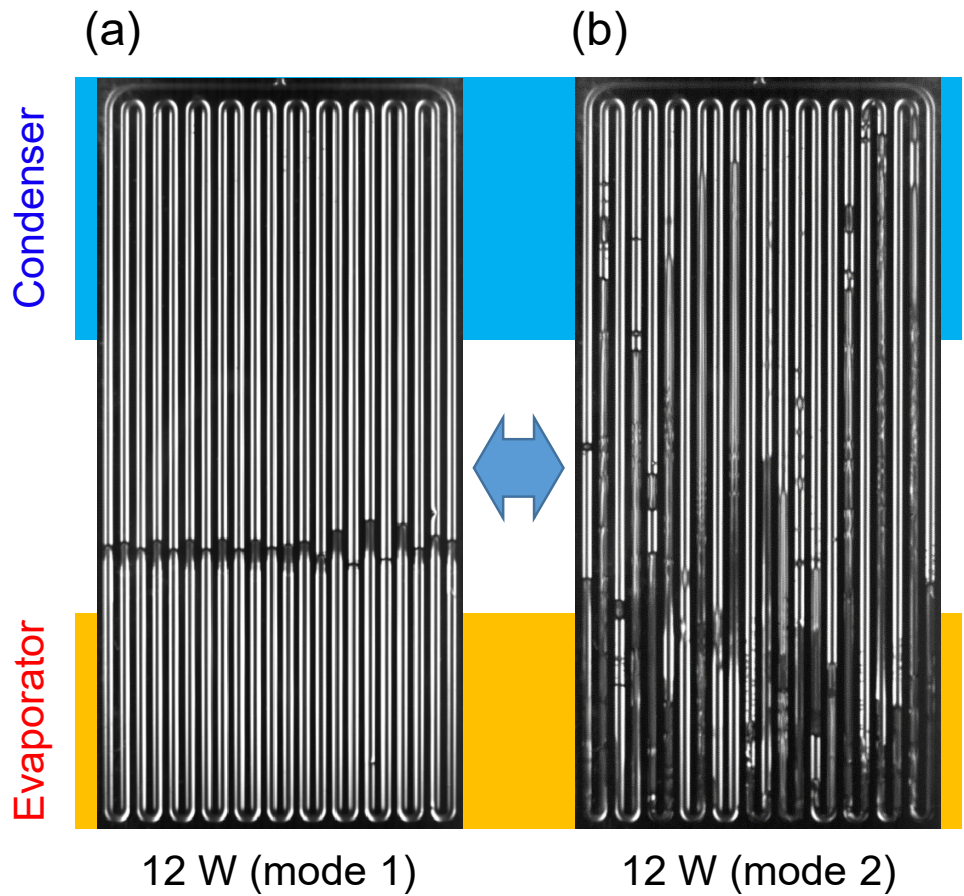

**Fig. 6.** Images of the internal flows for $\varphi$ = 63% at $T_c$ = 20 °C. The effective thermal conductivities for (a) $\dot{Q}_H$ = 12 W (mode 1) and (b) $\dot{Q}_H$ = 12 W (mode 2) are shown in Fig. 4(c). The working fluid self-oscillates for (b) whereas it almost halts for (a).

*4.2. Extracting flow patterns using an image recognition technique*

The internal flows in the MPHP are quite complicated, including liquid slugs separating and merging with each other, generation and evaporation of liquid films, and rapid-boiling of the working fluid. As mentioned in Sec. 4.1, liquid slugs resemble vapor plugs surrounded by dry walls (see Fig. 5). One may think that they can be distinguished in reference to positional information; those present around the condenser are presumed to be liquid slugs. However, if the heat input rate is large enough to cause self-oscillations with a large amplitude, allowing liquid slugs to frequently penetrate into the evaporator section, this assumption is no longer reasonable. In addition, when a liquid slug directly contacts a vapor plug surrounded by dry walls without a liquid film between them, the meniscus becomes too ambiguous to determine an accurate position.

It is necessary to automatically extract temporal changes in the positions of liquid slugs, liquid films, dry walls, and rapid-boiling regions for further analysis, since the number of pictures obtained in each experiment runs up to several hundreds of thousands. However, the internal flows are quite complicated, as explained above. It is difficult to address this issue with conventional image-processing algorithms such as edge detection or image thresholding; therefore, semantic segmentation [23], a deep-learning-based image recognition technique, is applied to the present task. Given an input picture, semantic segmentation automatically labels each pixel of the picture based on the type of object, for example, cars, buildings, trees, humans, etc. A conditional random field has been used as an algorithm to enable semantic segmentation [23,24], but some studies in recent years [25–27] have reported that applying convolutional neural networks (CNNs) to semantic segmentation remarkably improves accuracy. Hence, CNN-based semantic segmentation is employed for extracting the internal flow patterns of the MPHP.

*4.3. Construction of the image recognition model*

DeepLabv3+ [27], which achieves state-of-the-art accuracy, is selected from among the CNN architectures for semantic segmentation. This architecture also has the virtue of a low calculation cost in training and predicting. The original network has enormous parameters and complexity to handle difficult tasks such as segmentation of movies recorded by in-car cameras. Extracting the internal flow patterns is a simpler problem, thus the number of parameters is decreased to prevent overfitting and to reduce computational cost while maintaining a basic architecture. Fifty-five pictures of internal flows recorded with the high-speed camera are selected from various experimental conditions, and corresponding labels are assigned manually. The labels include four different flow patterns: liquid slugs, liquid films, rapid-boiling regions of the working fluid, and dry walls (Fig. 5). It is difficult to extract flow patterns at the corner sections of the MPHP illustrated in Fig. 7(a) because they are not illuminated uniformly. Thus, they are excluded from the image recognition process. Each picture includes 22 parallel straight channels and the total number of straight channels in the dataset is 1,210, which is sufficient for constructing the image recognition model. The dataset is split into 45 training datapoints, 5 validation datapoints, and 5 test datapoints. The accuracy of the trained model for the test dataset, defined as the proportion of correctly predicted pixels to the whole number of pixels, reaches 95.9%.

For further analysis and discussion, a one-dimensional meandering axis along the channel ($x$ axis) is introduced, as shown in Fig. 7(a). As an example, the image recognition results for 4 out of the 22 straight channels are provided in Fig. 7(b), where each position $x$ is tagged with one of the four flow patterns at each time step to generate time-series data. Repeated inflows of liquid slugs into the evaporator section are clearly captured in the image recognition results. It is also observed that liquid films are generated due to moving liquid slugs and gradually evaporate in the evaporator.

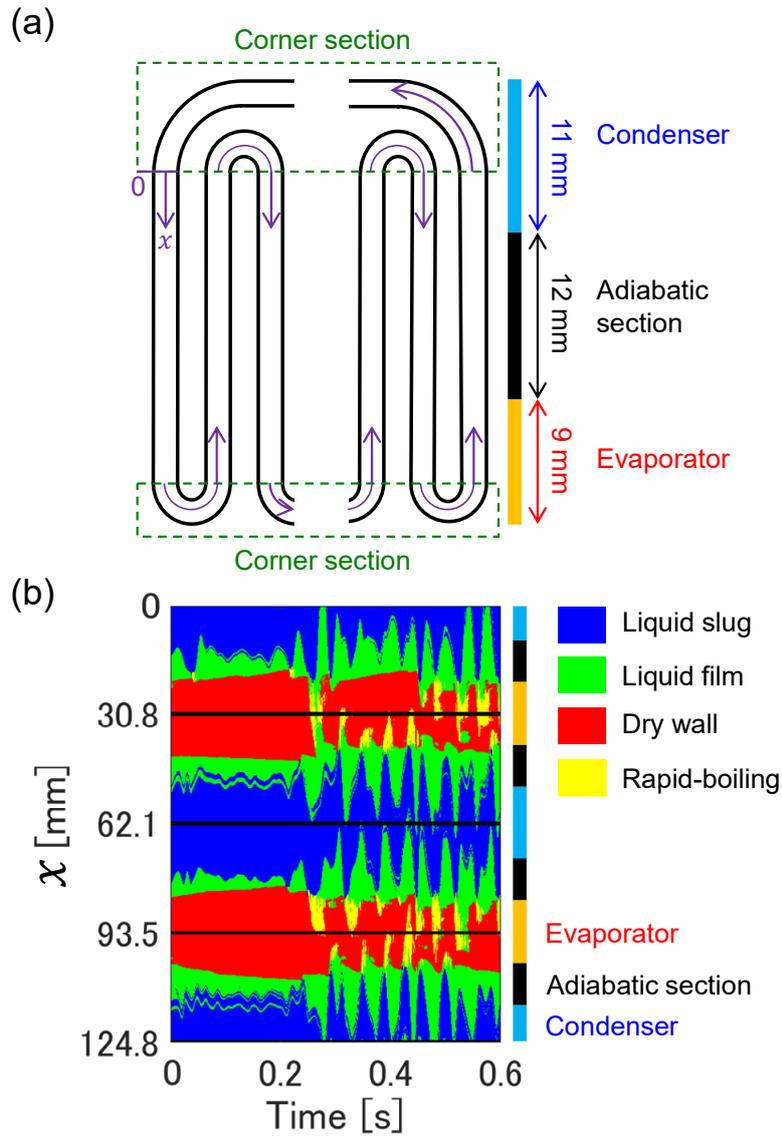

**Fig. 7.** (a) A meandering coordinate along the channel ($x$ axis). The corner edge, which connects the left side and the top side of the channel, is set to be the origin. Corner sections are excluded from the image recognition process. Dimensions are not to scale. (b) An example of the image recognition results for 4 out of the 22 straight channels ($T_c$ = 40 °C, $\varphi$ = 48%, and $\dot{Q}_H$ = 16 W). Each position $x$ is tagged with one of the four flow patterns at each time step to generate time-series data. Blue, green, red, and yellow regions correspond to liquid slugs, liquid films on the channel wall, dry walls without liquid films, and rapid-boiling, respectively. Black horizontal lines at $x$ = 30.8 mm, 62.1 mm, 93.5 mm, and 124.8 mm correspond to the corner sections.

*4.4. Approximation of the extracted flow patterns*

The liquid slugs oscillate along the channels while separating and merging with each other. For simplicity, we introduce a method for approximating those flow patterns extracted by image recognition. According to Fig 7(b), liquid slugs existing in each section between two adjacent corners in the evaporator can be approximated as a single continuous liquid slug, since in general, the liquid slugs move synchronously with a small fraction of the liquid film between them. Therefore, it is assumed that there exists only one long liquid slug in each section, which is shown in Fig. 8(a). In each section, the total length of multiple liquid slugs, $L_i$, and their center-of-mass, $x_{g,i}$, are calculated and the liquid slugs are replaced by one liquid slug whose length and center-of-mass are $L_i$ and $x_{g,i}$. We assume that each corner section omitted from the image recognition process is totally filled with a liquid slug if both pixels just before and after the corner section are labeled as such. We note that the liquid slugs do not flow out from each section, even if $\dot{Q}_H$ is large and intense self-oscillations occur. Since the MPHP has 11 turns, there are 11 liquid slugs corresponding to the 11 sections shown in Fig. 8(b).

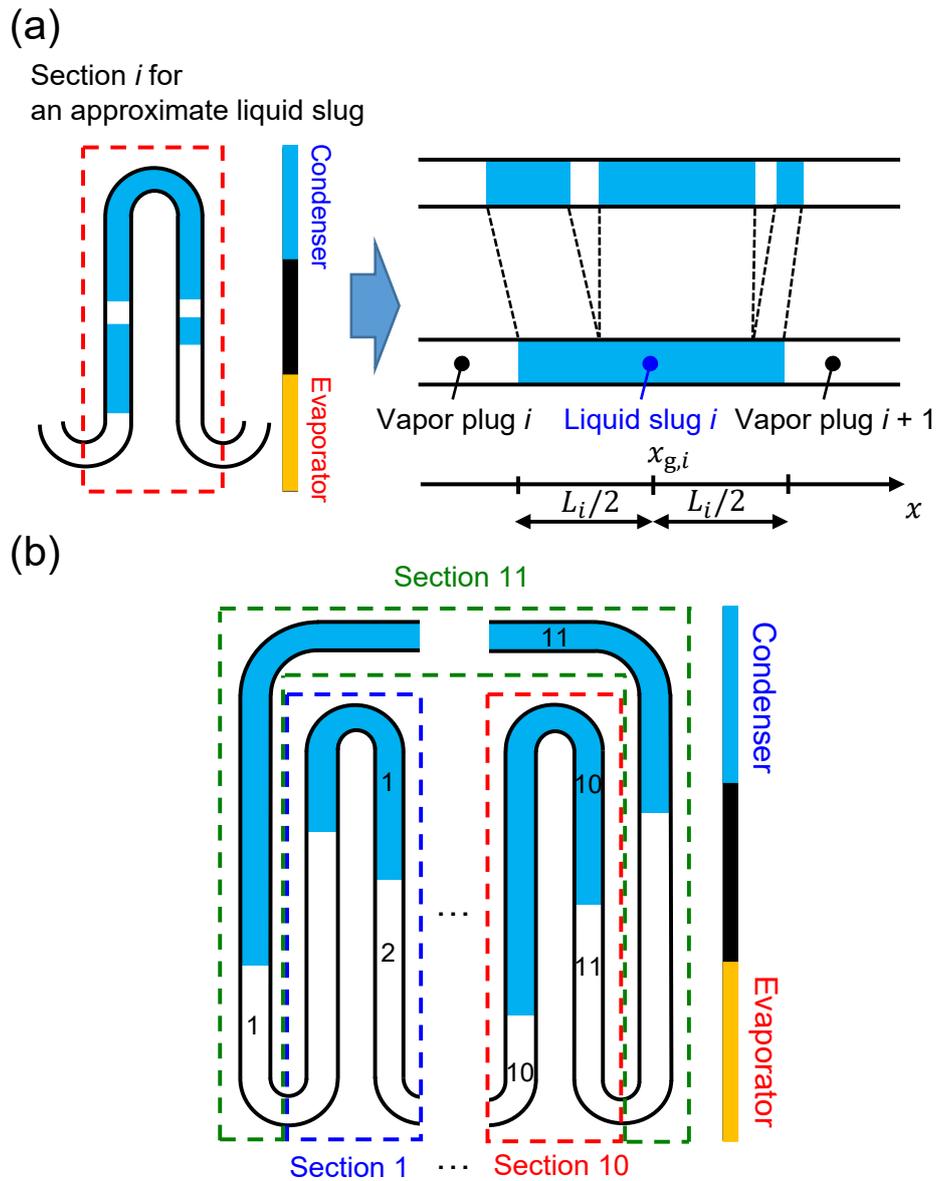

**Fig. 8.** Approximation of the original flow patterns extracted by image recognition. (a) Multiple liquid slugs are approximated as a single liquid slug of length $L_i$ and center-of-mass $x_{g,i}$, and denoted as liquid slug $i$ ($i = 1, \ldots, 11$). The vapor plugs before and after liquid slug $i$ are denoted as vapor plugs $i$ and $i + 1$. (b) The MPHP consists of eleven sections. Note that the vapor plug after liquid slug 11 corresponds to vapor plug 1.

## 4.5. Relationship between thermal properties and flow patterns

The raw data of extracted flow patterns for $\varphi$ = 48% at $T_c$ = 40 °C is shown in Fig. 9, where representative parts are provided for clarity. The self-oscillations of the working fluid are intermittent and unstable for $\dot{Q}_H$ = 8 W (Fig. 9(a)); these oscillations almost stop around 1.4 s $\lesssim$ $t$ $\lesssim$ 1.7 s, 1.9 s $\lesssim$ $t$ $\lesssim$ 2.8 s, and 3.5 s $\lesssim$ $t$ $\lesssim$ 4.0 s. By contrast, the self-oscillations are stable and the oscillation amplitudes become larger for $\dot{Q}_H$ = 24 W (Fig. 9(c)). This flow behavior is beneficial for heat transfer because the liquid slugs continually flow into the evaporator, leading to longer liquid films and intensive boiling. The resultant enhancement of latent heat transfer is of great importance to achieve the high effective thermal conductivity over 700 W/(m·K) measured in this condition, as shown in Fig. 3(b). However, the oscillation amplitudes become relatively small around 0.4 s $\lesssim$ $t$ $\lesssim$ 0.6 s, 1.6 s $\lesssim$ $t$ $\lesssim$ 2.0 s, and 2.6 s $\lesssim$ $t$ $\lesssim$ 3.0 s. This type of transition between the large- and small-amplitude oscillation phases in the MPHPs was previously reported in [20]. Figure 10 shows the frequency spectra of the centers-of-mass of the approximate liquid slugs (Fig. 8) for $\varphi$ = 48% at $T_c$ = 40 °C. Sharp peaks appear around 20–30 Hz when $\dot{Q}_H$ becomes larger, corresponding to stable oscillations observed in Fig. 9(c). Additionally, we observe only oscillation flows in this study and the MPHP never undergoes circulation flows which can further enhance heat transfer [7]. In the previous study [11], circulation flows were also observed in an MPHP with a hydraulic diameter of 394 μm, whereas only oscillation flows were observed in MPHPs with hydraulic diameters 352 μm or less, which is consistent with our results. We infer that an increase in the wall friction force with decreasing hydraulic diameter suppresses circulation flows.

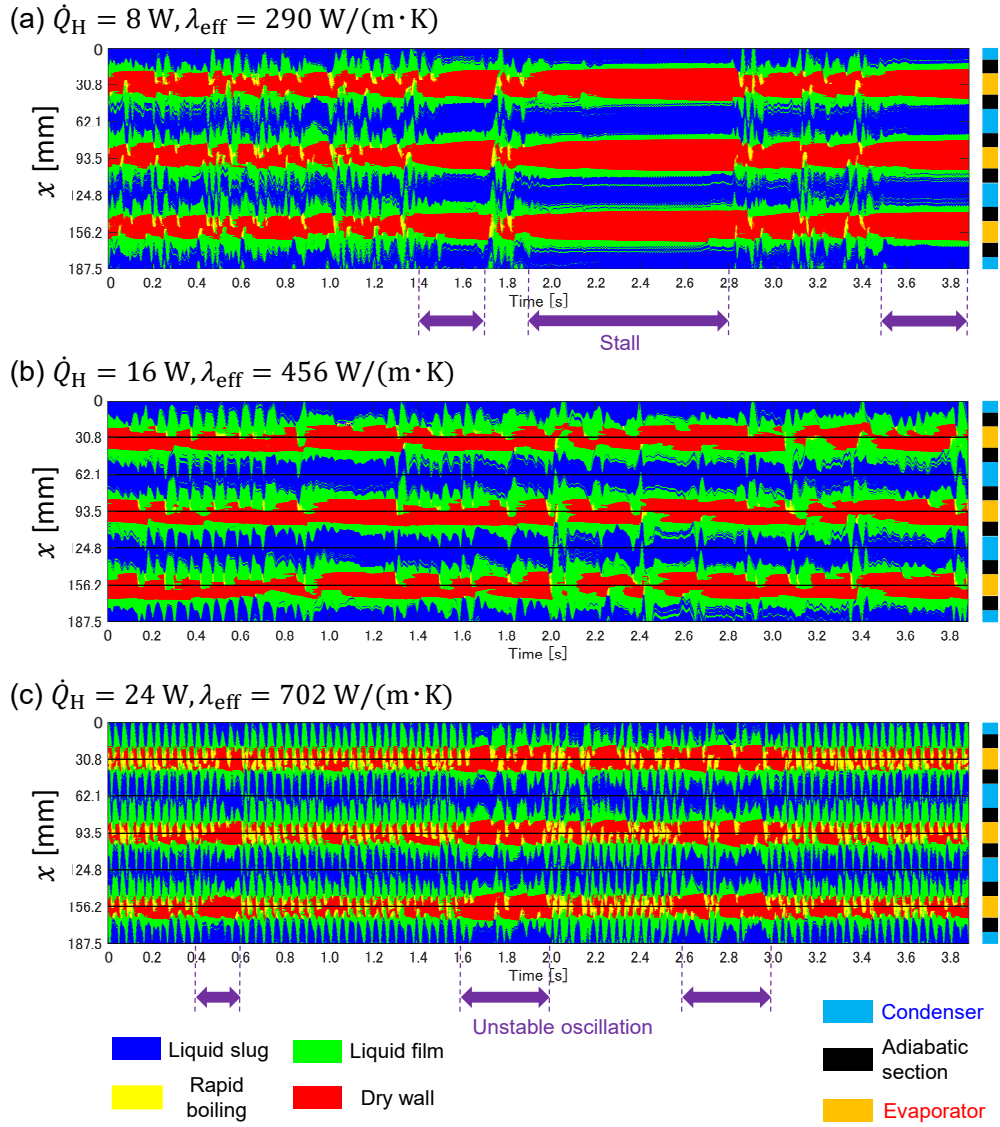

**Fig. 9.** Representative parts of the extracted flow patterns for $\varphi = 48\%$ at $T_c = 40$ °C. (a) $\dot{Q}_H = 8$ W, (b) $\dot{Q}_H = 16$ W, and (c) $\dot{Q}_H = 24$ W.

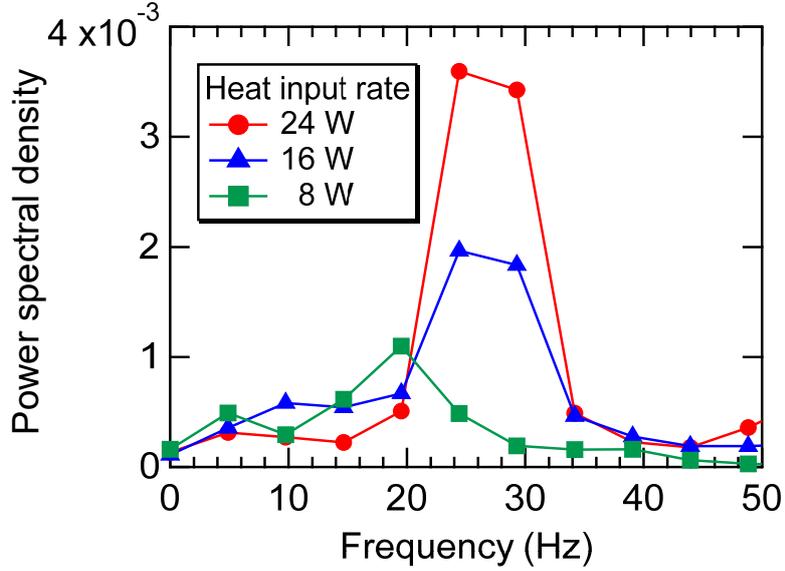

**Fig. 10.** Frequency spectra of the centers-of-mass of the approximate liquid slugs (Fig. 8) for $\dot{Q}_\mathrm{H}$ = 8 W, 16 W, and 24 W under $\varphi$ = 48% at $T_\mathrm{c}$ = 40 °C. The sampling frequency is 5 kHz, equivalent to the framerate of recording with the high-speed camera.

We calculate the averaged proportions of liquid-film and rapid-boiling regions in the flow channels, since these flow patterns are of great importance for latent heat transfer. The average thickness of the liquid films generated by moving liquid slugs is also estimated by substituting the velocities and accelerations of the approximate liquid slugs into the experimental-correlation equation [28]. The effective thermal conductivities, averaged proportions of liquid-film and rapid-boiling regions, and average liquid-film thickness for $\varphi$ = 48% at $T_\mathrm{c}$ = 40 °C, are summarized in Figs. 11(a)–(c). Under this experimental condition, the largest effective thermal conductivity of ≈ 700 W/(m·K) is measured in the present study. As shown in Fig. 11(b), the proportions of liquid-film and rapid-boiling regions tend to increase with $\dot{Q}_\mathrm{H}$. The large oscillation amplitudes of moving liquid slugs are responsible for increasing the film length. In addition, the liquid slugs repeatedly wet channel walls in the evaporator before the liquid films completely dry out, thanks to the stable high-frequency oscillations shown in Fig. 10. The larger proportions of liquid-film and rapid-boiling regions result in extended sections in which phase change occurs, thus enhancing latent heat transfer. Fig. 11(c) shows that liquid films gradually become thinner with increasing $\dot{Q}_\mathrm{H}$. Thinner liquid films are more likely to evaporate and thus lead to more efficient latent heat transfer. Therefore, we infer that an increase in latent heat transfer due to long and thin liquid films contributes to the high thermal conductivity over 700 W/(m·K). We also calculate the averaged proportions of liquid-film and rapid-boiling regions, together with the averaged liquid-film thickness for $\varphi = 39\%$ at $T_\mathrm{c} = 40$ ˚C, in which we observe remarkable dryout. The results are shown in Fig. 11(d)–(f), where the onset of dryout is associated with a rapid increase in the evaporator temperature $T_\mathrm{e}$ (from 105 ˚C to 145 ˚C). According to Fig. 11(e), rapid-boiling regions tend to increase with $\dot{Q}_\mathrm{H}$ when no dryout occurs. However, such regions significantly shrink under the dryout condition, meaning that the self-oscillation amplitude becomes smaller and liquid slugs do not flow into the evaporator sections. In contrast to $\varphi = 48\%$ at $T_\mathrm{c} = 40$ ˚C, liquid-film regions tend to decrease with increasing $\dot{Q}_\mathrm{H}$. This explains why dryout does not occur for $\varphi = 48\%$ at $T_\mathrm{c} = 40$ ˚C, but occurs for $\varphi = 39\%$ at $T_\mathrm{c} = 40$ ˚C. For $\varphi = 48\%$ at $T_\mathrm{c} = 40$ ˚C, liquid films become longer and efficient heat transfer is achieved when $\dot{Q}_\mathrm{H}$ increases, preventing occurrence of dryout. For $\varphi = 39\%$ at $T_\mathrm{c} = 40$ ˚C, flow behavior becomes less advantageous for heat transfer when $\dot{Q}_\mathrm{H}$ increases because of shorter liquid films. Once the thermal balance between the input heat $\dot{Q}_\mathrm{H}$ and the latent heat transferred by evaporation of liquid films collapses, the evaporator temperature drastically increases and finally dryout occurs. Furthermore, as shown in Fig. 11(f), liquid films become thicker when $\dot{Q}_\mathrm{H}$ increases from 14 W (without dryout) to 16 W (dryout). This is one cause of the low effective thermal conductivity under the dryout condition, because thicker liquid films are less likely to evaporate rapidly.

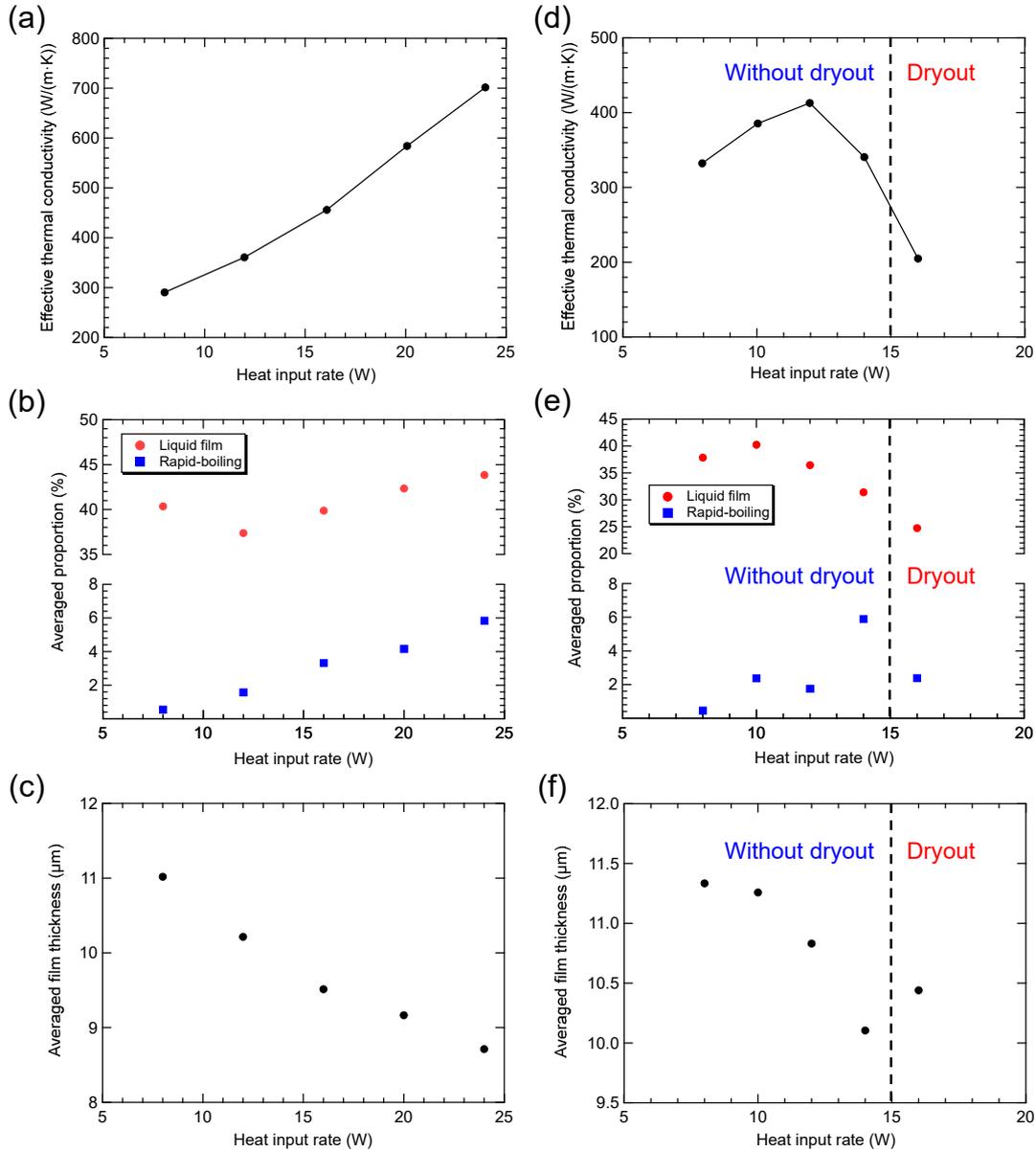

**Fig. 11.** The relationship between effective thermal conductivities and flow characteristics for $\varphi$ = 48% at $T_c$ = 40 °C ((a)–(c)) and $\varphi$ = 39% at $T_c$ = 40 °C ((d)–(f)). (a), (d) The effective thermal conductivities. (b), (e) Averaged proportions of liquid-film and rapid-boiling regions to the flow channels calculated from the extracted flow patterns. (c), (f) Averaged liquid-film thickness estimated using the velocities and accelerations of approximate liquid slugs (Fig. 8). The onset of dryout is associated with a rapid increase in the evaporator temperature $T_e$ (from 105 °C to 145 °C).

## 5. Modeling and analysis of internal flows and heat transfer of the MPHP

To analyze heat transfer properties, we construct a numerical model of the MPHP based on previous studies [13,16,29], where the approximate flow patterns (Sec. 4.4) and experimentally measured temperatures are incorporated as inputs. The model assumes slug flow including liquid slugs and vapor plugs, which are observed via the flow visualization illustrated in Fig. 5. The present model does not take rapid-boiling regions into account due to the difficulty of simulating boiling phenomena; rather, they are treated in the same way as liquid films. Here, heat transfer in the MPHP is divided into two factors: sensible/latent heat transfer via vapor plugs (Sec. 5.1) and sensible heat transfer via liquid slugs (Sec. 5.2), as illustrated in Fig. 12. Heat transfer via vapor plugs involves sensible heat transfer between the vapor phase and the dry walls or liquid films, and latent heat transfer due to the phase change of liquid films is important for heat transfer in microchannels [30]. Sensible heat transfer via liquid slugs is caused by the temperature distributions in liquid slugs due to heat flux from the channel walls. The flow patterns and wall temperatures obtained for $\varphi$ = 39% and 48% at $T_c$ = 40 ˚C are used for the subsequent numerical analyses.

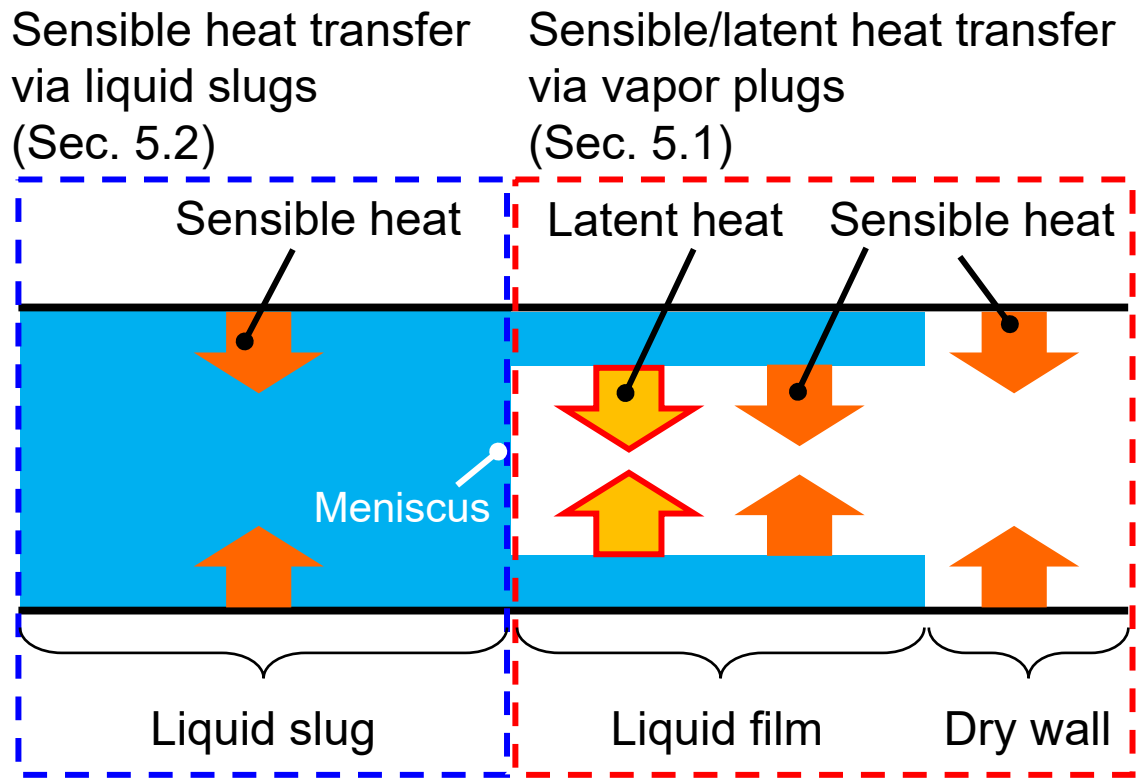

**Fig. 12.** Heat transfer considered in the present model: sensible heat transfer via liquid slugs and sensible/latent heat transfer via vapor plugs. The latter involves sensible heat transfer between the vapor phase and dry walls or liquid films, as well as latent heat transfer due to the phase change of liquid films.

## 5.1. Sensible/latent heat transfer via vapor plugs

The following assumptions are made to model sensible/latent heat transfer via vapor plugs:

(a) Latent heat transfer via liquid films is considered while that via meniscuses is neglected because meniscus areas are much smaller than liquid-film areas.

(b) Heat transfer through liquid films is modeled as one-dimensional conductive heat transfer in the transverse direction, as the liquid films are thin.

(c) Heat transfer between the channel wall/liquid films and the vapor phase is modeled as one-dimensional heat transfer in the transverse direction, while the vapor phase heat transfer in the flow direction is neglected.

(d) Liquid surface temperature is equal to the saturation temperature of a vapor plug.

Modeling of heat transfer via vapor plugs is conducted for each vapor plug (Fig. 8). A schematic diagram of the model is illustrated in Fig. 13. The equation of state for vapor plug $i$ is given as

$$P_i V_i = n_i z_i R T_{v,i} \qquad (6)$$

where $R$, $P_i$, $V_i$, $n_i$, $z_i$, and $T_{v,i}$ are the universal gas constant, pressure, volume, molar number, compressibility factor, and temperature of vapor plug $i$, respectively. The energy equation for vapor plug $i$ is expressed as

$$\dot{U}_i = \dot{Q}_{\text{sen},i} + \dot{H}_{\text{lat},i} - P_i \dot{V}_i \qquad (7)$$

where $U_i$, $\dot{Q}_{\text{sen},i}$, and $\dot{H}_{\text{lat},i}$ are the internal energy, sensible heat transfer rate, and enthalpy inflow rate due to latent heat transfer, respectively. The sensible heat flux in the transverse direction between the vapor phase and the channel wall or liquid film, $q''_{\text{sen}}$, can be estimated by

$$q''_{\text{sen}} = \frac{\lambda_v}{a/2} (T_{\text{sur}} - T_{v,i}) \qquad (8)$$

where $\lambda_v$, $a$, and $T_{\text{sur}}$ are the thermal conductivity of vapor, the length of one side of the channel (350 µm), and the surface temperature of the channel wall or liquid film, respectively. $T_{\text{sur}}$ is given as

$$T_{\text{sur}} = \begin{cases} T_w & \text{(on channel walls)} \\ T_{\text{sat}}(P_i) & \text{(on liquid films)} \end{cases} \qquad (9)$$

where $T_\text{w}$ and $T_\text{sat}(P_i)$ correspond, respectively, to the wall and saturation temperatures of vapor plug $i$. Note that the surface temperature of the liquid film is equal to $T_\text{sat}(P_i)$ according to assumption (d). Then, $\dot{Q}_{\text{sen},i}$ in Eq. (7) is obtained by

$$\dot{Q}_{\text{sen},i} = 4a \int_{\text{v},i} q''_\text{sen} \, dx \tag{10}$$

where "v, $i$" indicates the integral along the $x$ axis within vapor plug $i$, shown in Fig. 13(a). From assumptions (b) and (d), the heat flux through the liquid film in the transverse direction, $q''_\text{film}$, is calculated as

$$q''_\text{film} = \frac{\lambda_\text{l}(T_\text{w} - T_\text{sat}(P_i))}{\delta} \tag{11}$$

where $\delta$ and $\lambda_\text{l}$ indicate the liquid-film thickness and thermal conductivity of the liquid. Since the liquid-film thickness is estimated to be very small ($\delta \sim 10$ μm; see Fig. 11(c)), a linear temperature distribution between the liquid-film surface and the channel wall is assumed in the transverse direction, as illustrated in Fig. 13(c). Note that although Eq. (6) assumes a homogenous temperature of the vapor phase, $T_{\text{v},i}$, sensible heat transfer between the vapor phase and channel wall or liquid film is taken into account via Eq. (8). Furthermore, non-uniform liquid-film thickness is considered in this model and $\delta$ depends on the $x$ axis. Considering the energy balance of the phase change on the liquid-film surface, the following equation is obtained:

$$q''_\text{lat} = m'' h_\text{lv} = q''_\text{film} - q''_\text{sen} \tag{12}$$

where $q''_\text{lat}$, $m''$, and $h_\text{lv}$ correspond to the latent heat flux, mass flux, and specific heat of vaporization on the liquid-film surface, respectively. $m'' > 0$ and $m'' < 0$ correspond to evaporation and condensation. Then, $\dot{H}_{\text{lat}.i}$ in Eq. (7) is calculated as

$$\dot{H}_{\text{lat}.i} = 4a \int_{\text{film},i} m'' h_\text{v} dx \tag{13}$$

where "film, $i$" and $h_\text{v}$ indicate the liquid film in vapor plug $i$ and the specific vapor enthalpy on the liquid film surface, respectively.

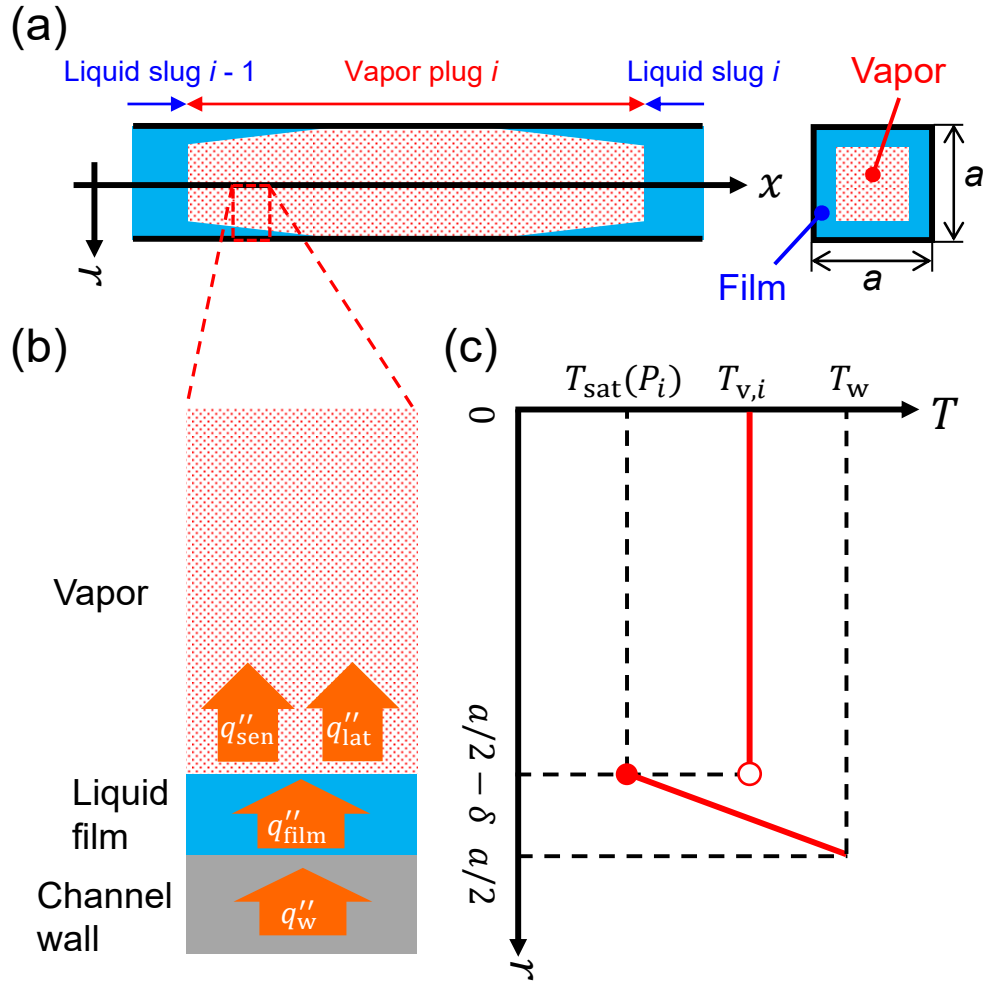

**Fig. 13.** A heat transfer model for vapor plugs. (a) Definitions of an axial coordinate ($x$-axis) and a transverse coordinate ($r$-axis). (b) An enlarged view of the region comprising the vapor phase, liquid film, and channel wall denoted by the dashed line in (a). In the presence of the liquid film, the energy balance $q''_w = q''_{film} = q''_{sen} + q''_{lat}$ holds, where $q''_w$ is the heat flux through the channel wall, $q''_{film}$ is that through the liquid film, $q''_{sen}$ is the sensible heat flux from the liquid film to the vapor, and $q''_{lat}$ is latent heat flux from the liquid film to the vapor, respectively. (c) A schematic of the temperature distribution through the vapor phase and the liquid film corresponding to (b). Although Eq. (6) assumes a homogeneous temperature of the vapor phase, $T_{v,i}$, sensible heat transfer between the vapor phase and the liquid film, $q''_{sen}$, is taken into account via Eq. (8).

Numerical analysis is conducted by substituting the meniscus positions, liquid-film distributions, and temperature distributions obtained from the experiments into the model. The meniscus positions at each time step are obtained from the 11 approximate liquid slugs (Fig. 8). The distribution of the wall temperature is estimated by linear interpolation/extrapolation using the temperatures measured by the four thermocouples shown in Fig. 2(c). The positions of liquid films and the distributions of liquid-film thickness along the flow channel are also estimated from the extracted flow patterns. At each time step, we compare the positions of the liquid films with those in the previous step. If liquid films exist in the regions in which no liquid films existed at the previous step, they are considered to be generated by the liquid slugs moving between the two steps, and their thicknesses are calculated by substituting the velocities and accelerations of the nearest liquid slugs into the experimental-correlation equation [28]. The physical properties of the gas phase are estimated using a four-parameter equation of state [31], whose parameters are obtained from a previous study [32]. The internal energy at an ideal-gas state is required to calculate the enthalpy at a real-gas state using the four-parameter equation of state. To this end, we calculate the molecular vibrations of FC-72 using the Gaussian 16 electronic structure package [33] with B3LYP/6-31G(d) level of theory with a frequency scaling factor of 0.961 [34], and estimate the internal energy from the vibrational frequencies. The temperature-dependent physical properties of the liquid phase are obtained from previous datasheets [35,36]. Since liquid films have temperature distributions as shown in Fig. 13(c), temperature-dependent $\lambda_l$ in Eq. (11) is calculated from the temperature averaged along the $r$-axis, $(T_w + T_{sat}(P_i))/2$, for each liquid film at each time step. At each time step, $V_i$ is estimated from the meniscus position and the mass change rate is obtained from Eqs. (8), (11), and (12). After calculating $\dot{Q}_{sen,i}$ and $\dot{H}_{lat,i}$ from Eqs. (10) and (13), we update $P_i$ and $T_{v,i}$ using Eqs. (6) and (7).

The heat transfer rate to the condenser via vapor plug $i$, $\dot{Q}_{v,i}^{cnd}$, is calculated by

$$\dot{Q}_{v,i}^{cnd} = 4a \int_{film,i,cnd} q_{film}'' dx + 4a \int_{dry,i,cnd} q_{sen}'' dx \qquad (14)$$

where "film, $i$, cnd" and "dry, $i$, cnd" correspond to the liquid film and dry wall of vapor plug $i$ located in the condenser. The overall heat transfer rate to the condenser via all vapor plugs, $\dot{Q}_v^{cnd}$, is given as

$$\dot{Q}_v^{cnd} = \sum_i \dot{Q}_{v,i}^{cnd} \qquad (15)$$

The cumulative heat transfer to the condenser via all vapor plugs, $Q_v^{cnd}(t) = \int_0^t \dot{Q}_v^{cnd} dt$, is shown in Fig. 14 for $\varphi = 48\%$, $T_c = 40$ °C, and $\dot{Q}_H = 24$ W. Although $Q_v^{cnd}$ almost linearly increases with time, this rate of increase becomes smaller around $t = 0.5$ s, 1.8 s, and 2.8 s. These correspond to unstable oscillation modes observed in Fig. 9(c), suggesting that the existence of unstable small-amplitude oscillations deteriorates the thermal performance of the MPHP.

The heat transfer rate to the condenser via the channel wall, $\dot{Q}_{wall}^{cnd}$, is also estimated as

$$\dot{Q}_{wall}^{cnd} = \lambda_{si} S_{si} \frac{T_1 - T_2}{d} \qquad (16)$$

where $T_1$ and $T_2$ correspond to the temperatures measured by the thermocouples shown in Fig. 2(c), $\lambda_{si}$ and $S_{si} = 7.805$ mm² are the temperature-dependent thermal conductivity [37] and the cross-sectional area of the silicon substrate, and $d = 8$ mm is the distance between thermocouples 1 and 2. The thermal conductivity of the glass plate is much smaller than that of the silicon wall and is therefore neglected. In addition, we decompose $\dot{Q}_v^{cnd}$ into the latent heat transfer rate from the liquid films and the sensible heat transfer rate from the channel wall/liquid films, as shown in Fig. 15. The calculated heat transfer rates show reasonable agreement with the heat input rates, although the present model has not yet taken into account sensible heat transfer via liquid slugs, which will prove to be much less significant in Sec. 5.2. More importantly, the latent heat transfer via liquid films accounts for a considerable portion of the total heat transfer rate, while the sensible heat transfer contribution is negligible. Additionally, the calculated heat transfer rate still agrees well with the heat input rate, even under the dryout condition ($\dot{Q}_H = 16$ W), as shown in Fig. 15(b). Although $\dot{Q}_v^{cnd}$ is smaller for $\dot{Q}_H = 16$ W due to the dryout compared to $\dot{Q}_H = 14$ W, the resultant higher evaporator temperature increases $\dot{Q}_{wall}^{cnd}$ and hence the total heat transfer rate $\dot{Q}_v^{cnd} + \dot{Q}_{wall}^{cnd}$ remains close to the heat input rate.

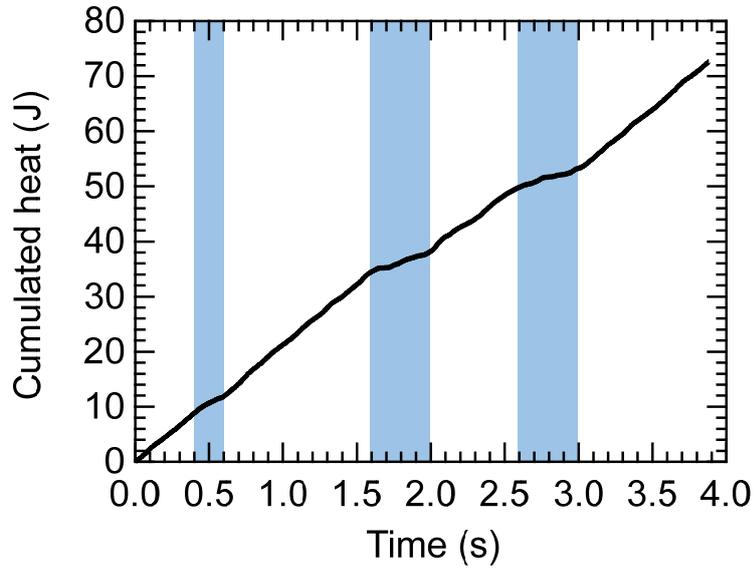

**Fig. 14.** The cumulative heat transfer to the condenser via all vapor plugs, $Q_v^{cnd}(t)$, for $\varphi = 48\%$, $T_c = 40$ °C, and $\dot{Q}_H = 24$ W. Blue-shaded areas around 0.4 s $\lesssim t \lesssim$ 0.6 s, 1.6 s $\lesssim t \lesssim$ 2.0 s, and 2.6 s $\lesssim t \lesssim$ 3.0 s correspond to the unstable oscillations observed in Fig. 9(c).

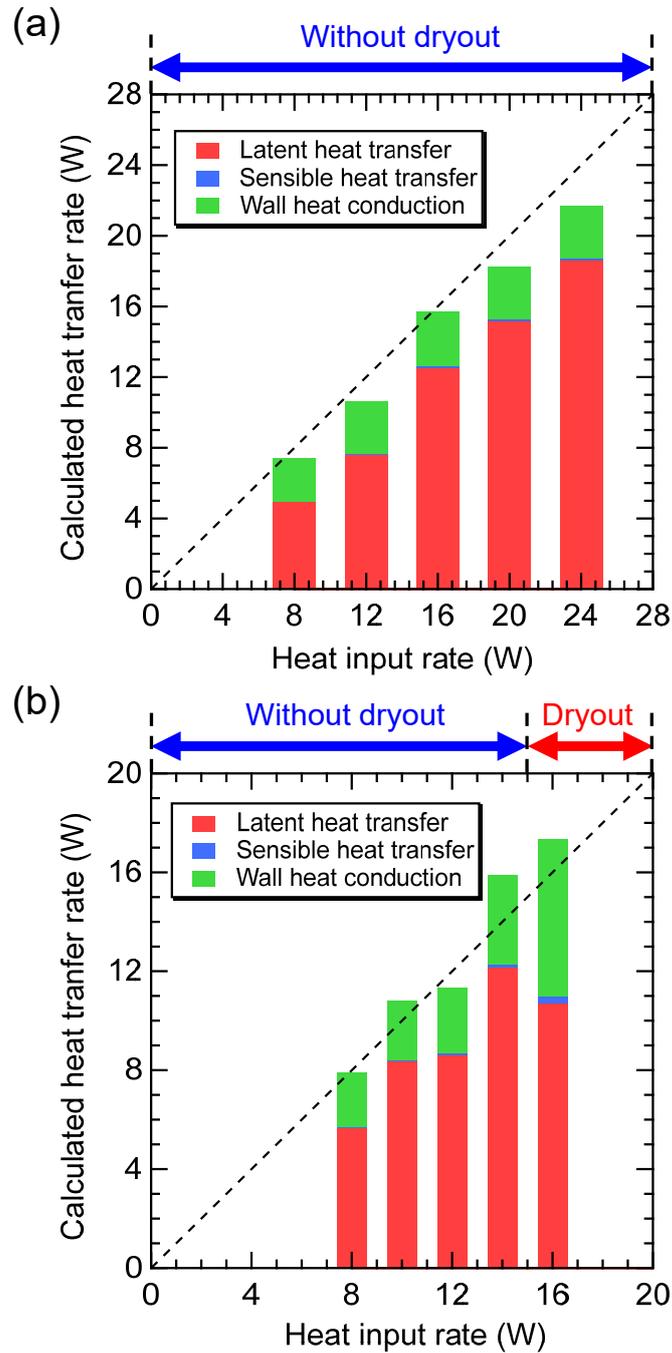

**Fig. 15.** Calculated heat transfer rates to the condenser based on latent heat transfer from liquid films to vapor plugs, sensible heat transfer from the channel wall and liquid films to the vapor plugs, and heat conduction via the channel wall, $\dot{Q}_{\text{wall}}^{\text{cnd}}$: (a) $\varphi = 48\%$ at $T_c = 40$ °C and (b) $\varphi = 39\%$ at $T_c = 40$ °C. Dashed lines are provided for comparison with the heat input rates.

## 5.2. Sensible heat transfer via liquid slugs

A one-dimensional heat transfer model is considered to evaluate sensible heat transfer via liquid slugs (Fig. 12). The heat transfer rate is estimated for each approximate liquid slug (Fig. 8), and then the overall heat transfer rate by all the liquid slugs is calculated by adding them up. For liquid slug $i$, the coordinate fixed at the liquid slug in the flow direction, $X_i$, is introduced. The two meniscuses of liquid slug $i$ correspond to $X_i = 0$ and $L_i$, where $L_i$ is the length of liquid slug $i$ (Fig. 8). The thermal diffusion equation for liquid slug $i$ is given as

$$\rho_l c_{p,l} a^2 \frac{\partial T_{l,i}}{\partial t} = \lambda_l a^2 \frac{\partial^2 T_{l,i}}{\partial X_i^2} + 4aq_w'' \tag{17}$$

where $\rho_l$, $c_{p,l}$, $\lambda_l$, and $T_{l,i}$ denote the density, specific heat, thermal conductivity, and temperature of liquid slug $i$, and $q_w''$ is heat flux from the wall. Assuming an isothermal wall in the evaporator and condenser, $q_w''$ is given as

$$q_w'' = \begin{cases} \dfrac{\lambda_l Nu}{a}(T_H - T_{l,i}) & \text{(Evaporator)} \\ 0 & \text{(Adiabatic section)} \\ \dfrac{\lambda_l Nu}{a}(T_L - T_{l,i}) & \text{(Condenser)} \end{cases} \tag{18}$$

where $Nu$ is the Nusselt number and $T_H$ and $T_L$ are temperatures at the evaporator and condenser measured by the thermocouples shown in Fig. 2(c). The boundary conditions on the meniscuses are given as

$$\begin{aligned} T_{l,i}(X_i = 0) &= T_{sat}(P_i) \\ T_{l,i}(X_i = L_i) &= T_{sat}(P_{i+1}) \end{aligned} \tag{19}$$

where $T_{sat}(P_i)$ and $T_{sat}(P_{i+1})$ are the saturation temperatures of the two adjacent vapor plugs (Fig. 8(a)). If $L_i$ increases compared to the previous time step, the temperatures of the incremental liquids on both sides of liquid slug $i$ are set to be the same as $T_{sat}(P_i)$ and $T_{sat}(P_{i+1})$.

The heat transfer rate to the condenser via liquid slug $i$, $\dot{Q}_{l,i}^{cnd}$, is defined by

$$\dot{Q}_{l,i}^{cnd} = 4a \int_{liquid,i,cnd} q_w'' dX_i \tag{20}$$

where "liquid, $i$, cnd" corresponds to part of liquid slug $i$ located in the condenser.

The heat transfer rate obtained from this model depends on the Nusselt number in Eq. (18).

However, it is difficult to obtain an accurate Nusselt number under unsteady oscillating flows and large temperature gradients along the channel. Therefore, we estimate upper and lower limits for the Nusselt number and estimate the heat transfer for several values of this parameter. For a Poiseuille flow developing in a square channel under the isothermal wall condition, the Nusselt number gradually becomes smaller as the temperature profile develops, finally approaching 2.89 [22]. Therefore, the lower limit of 2.89 is adopted. Additionally, the local Nusselt number $Nu_x$ for a thermally developing Poiseuille flow under the isothermal wall condition [22] is used to estimate the upper limit of the Nusselt number. Here, $Nu_x \sim 10$ is estimated using the liquid slug velocities and the self-oscillation amplitudes obtained from the extracted flow patterns. Therefore, the upper limit of the Nusselt number is set to 10. By substituting the positions of the liquid slugs and temperatures of the evaporator and condenser into this model at each time step, the heat transfer rate to the condenser via all liquid slugs, $\dot{Q}_l^{cnd} = \sum_i \dot{Q}_{l,i}^{cnd}$, is calculated. For each liquid slug, a temperature-dependent thermal conductivity is calculated by averaging the temperature of the liquid slug along the flow channel at each time step. The results are shown for $Nu$ = 2.89, 5, 7.5, and 10 in Fig. 16. Although $\dot{Q}_l^{cnd}$ increases with Nusselt number, its values are significantly smaller than $\dot{Q}_v^{cnd}$ shown in Fig. 15, regardless of whether dryout occurs. For example, the maximum $\dot{Q}_l^{cnd}$ for $\varphi$ = 48%, $T_c$ = 40 ˚C, and $\dot{Q}_H$=24 W is about 2.8 W, which is much smaller than $\dot{Q}_v^{cnd}$ = 18.6 W (Fig. 15(a)). These results indicate that sensible heat transfer via liquid slugs is much smaller than the heat transfer via vapor plugs, or more specifically, latent heat transfer via liquid films. Therefore, Figs. 15 and 16 indicate that the calculated total heat transfer rates, $\dot{Q}_v^{cnd} + \dot{Q}_{wall}^{cnd} + \dot{Q}_l^{cnd}$, show reasonable agreement with the experimental $\dot{Q}_H$ regardless of the Nusselt number affecting $\dot{Q}_l^{cnd}$, which ensures reliability of our numerical simulation results.

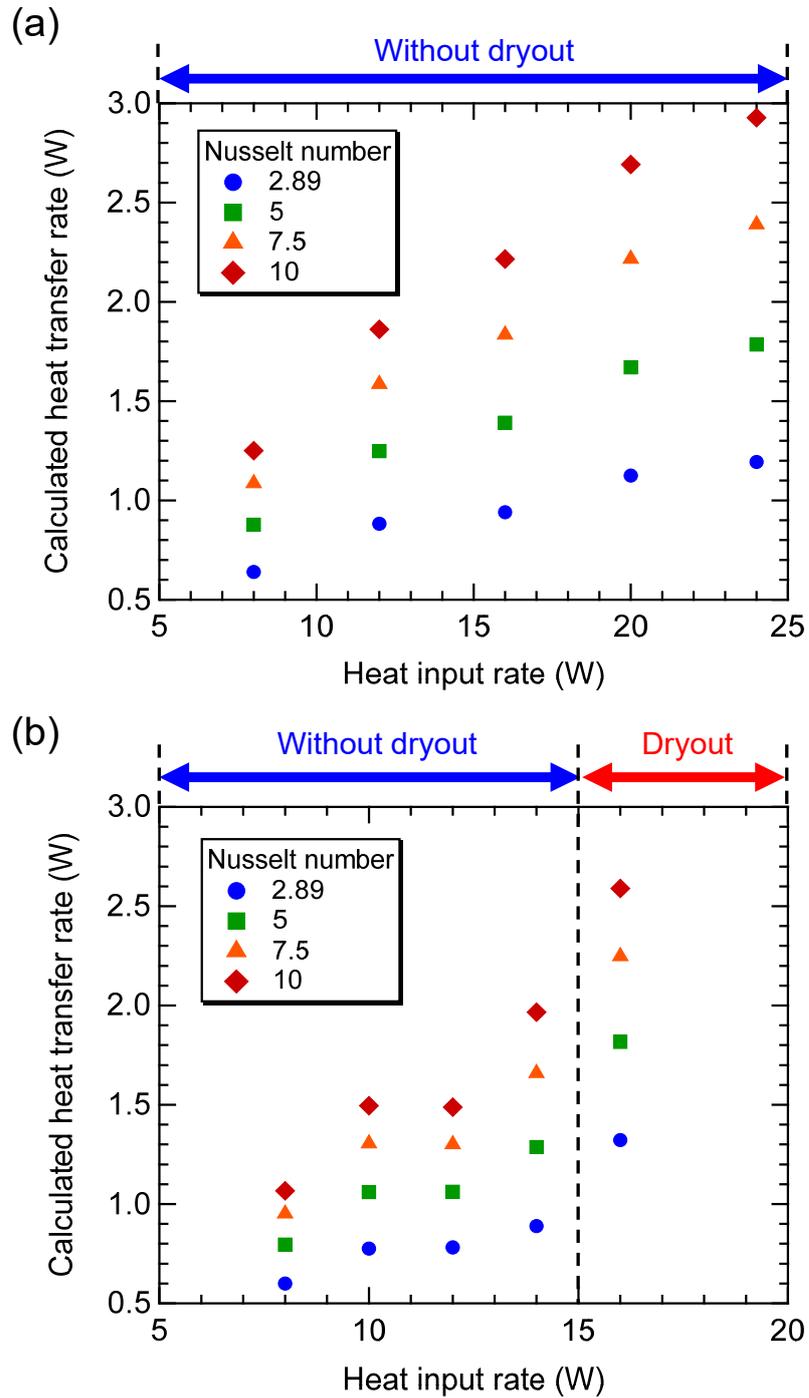

**Fig. 16.** Sensible heat transfer rate via liquid slugs, $\dot{Q}_l^{\mathrm{cnd}}$, calculated for $Nu$ = 2.89, 5, 7.5, and 10. (a) $\varphi$ = 48% at $T_c$ = 40 °C, where no dryout occurs for all the heat input rates. (b) $\varphi$ = 39% at $T_c$ = 40 °C, where dryout occurs for a heat input rate larger than 15 W.

6. Conclusions

In this study, we have investigated the relationship between the thermal properties of a micro pulsating heat pipe (MPHP) and its internal flow characteristics based on measurements of effective thermal conductivities, flow visualization followed by image recognition, and numerical simulations of heat transfer employing the extracted flow patterns and measured channel temperatures. The eleven-turn closed-loop MPHP is fabricated by dry-etching a meandering square microchannel with a hydraulic diameter of 350 μm onto a silicon substrate, which is then covered with a transparent glass plate to allow for flow imaging via a high-speed camera. The working fluid is degassed Fluorinert FC-72 with a filling ratio $\varphi$ varied from 39% to 63%, and the coolant temperature is set to be $T_c$ = 20 ˚C and 40 ˚C. Semantic segmentation-based image recognition successfully extracts four different flow patterns involving liquid slugs, liquid films, dry walls, and rapid-boiling regions. These extracted flow patterns in turn facilitate numerical simulations that take account of sensible/latent heat transfer via vapor plugs and sensible heat transfer via liquid slugs. Taken together, the experiments, the image recognition of the flow patterns, and the heat transfer simulations lead us to draw the following conclusions.

1. The effective thermal conductivities of the MPHP tend to improve for $T_c$ = 40 ˚C compared to $T_c$ = 20 ˚C. High effective thermal conductivities are measured for $\varphi \approx 50\%$. Specifically, the highest effective thermal conductivity of about 700 W/(m·K) is obtained for $\varphi$ = 48% and $T_c$ = 40 ˚C, with a heat input rate of $\dot{Q}_H$ = 24 W.
2. The MPHP exhibits two different self-oscillation modes with different thermal conductivities, even for the identical heat input rates. This tendency originates from the difference in the heat input rates at which the MPHP falls into and recovers from dryout. Once dryout occurs at a certain threshold of $\dot{Q}_H$, $\dot{Q}_H$ needs to be reduced below this threshold for the MPHP to recover from the dryout, which forms a hysteresis of the effective thermal conductivity.
3. The semantic segmentation-based image recognition of the flow patterns identifies stable self-oscillations with large amplitudes and high frequencies under the high thermal conductivity condition. Furthermore, longer and thinner liquid films are also estimated under these conditions, facilitating latent heat transfer via liquid films.
4. Heat transfer simulations incorporating the extracted flow patterns and the measured temperatures indicate that most transfer is of latent/sensible heat via vapor plugs, with sensible heat transfer via liquid slugs being less significant. More specifically, latent heat transfer via liquid films comprises a considerable portion of overall heat transfer.

## Acknowledgments

The authors would like to thank Prof. Yuji Suzuki, Dr. Minhyeok Lee, and Dr. Tomoya Miyoshi for letting us use their ultrasonic drill and anodic-bonding apparatus to fabricate the MPHP, together with their kind instructions.